\documentclass{svmult}

\usepackage{mathptmx}       
\usepackage{helvet}         
\usepackage{courier}        
\usepackage{type1cm}        
\usepackage{makeidx}         
\usepackage{graphicx}        
\usepackage{multicol}        
\usepackage[bottom]{footmisc}
\usepackage{amssymb}
\usepackage[numbers]{natbib}

\makeindex             
\newcommand{\apj}{ApJ}
\newcommand{\apjl}{ApJ}
\newcommand{\apjs}{ApJS}
\newcommand{\aap}{A\&A}
\newcommand{\aj}{AJ}
\newcommand{\mnras}{MNRAS}

\newcommand{\nat}{Nature}

\begin{document}
\title*{High-Redshift Galaxy Surveys and the Reionization of the Universe}
\author{Rychard Bouwens}
\institute{Rychard Bouwens \at Leiden Observatory, Niels Bohrweg 2, Leiden NL2333, Netherlands, \email{bouwens@strw.leidenuniv.nl}}
%
%
\maketitle \abstract*{Star-forming galaxies in the early universe
  provide us with perhaps the most natural way of explaining the
  reionization of the universe.  Current observational results are
  sufficiently comprehensive, as to allow us to approximately
  calculate how the ionizing radiation from galaxies varies as a
  function of cosmic time.  Important uncertainties in modeling
  reionization by galaxies revolve around the escape fraction and its
  luminosity and redshift dependence, a possible truncation of the
  galaxy luminosity function at the faint end, and an evolution in the
  production efficiency of Lyman-continuum photons with cosmic time.
  Despite these uncertainties, plausible choices for these parameters
  naturally predict a cosmic ionizing emissivity at $z\sim6$-10 whose
  evolution and overall normalization is in excellent agreement with
  that derived from current observational constraints.  This strongly
  suggests that galaxies provide the necessary photons to reionize the
  universe.}

\abstract{Star-forming galaxies in the early universe provide us with
  perhaps the most natural way of explaining the reionization of the
  universe.  Current observational results are sufficiently
  comprehensive, as to allow us to approximately calculate how the
  ionizing radiation from galaxies varies as a function of cosmic
  time.  Important uncertainties in modeling reionization by galaxies
  revolve around the escape fraction and its luminosity and redshift
  dependence, a possible truncation of the galaxy luminosity function
  at the faint end, and an evolution in the production efficiency of
  Lyman-continuum photons with cosmic time.  Despite these
  uncertainties, plausible choices for these parameters naturally
  predict a cosmic ionizing emissivity at $z\sim6$-10 whose evolution
  and overall normalization is in excellent agreement with that
  derived from current observational constraints.  This strongly
  suggests that galaxies provide the necessary photons to reionize the
  universe.}

\section{Introduction}
\label{sec:1}


One of the most important questions in observational cosmology regards
the reionization of the neutral hydrogen in the universe.  Over what
time scale did reionization occur and which sources caused it?
Observationally, we have constraints on reionization from the
Gunn-Peterson trough in luminous high-redshift quasars
\citep{XF02,XF06}, the Thomson optical depths observed in the
Microwave background radiation \citep{GH13}, and the luminosity
function and clustering properties of Ly$\alpha$ emitters
\citep[e.g.,][]{MO10}.  Reionization appears to have begun at least as
early as $z\sim11$ \citep{PL15}, with a midpoint at $z\sim9$, and
finished no later than $z\sim6$ \cite[e.g.,][]{MO10,XF06,IM11,IM15}.

Due to the low volume densities of QSOs at high redshift \citep[e.g.,
][]{CW10,IM13} and the lack of compelling evidence for other ionizing
sources (e.g., self-annihilating dark matter or mini-quasars:
\cite{MR04,PM04}), star-forming galaxies represent the most physically
well-motivated source of ionizing photons.  Observational surveys of
galaxies in the distant universe therefore provide us with our best
guide to mapping how the volume density of ionizing radiation likely
varies with both redshift and cosmic time.

Fortunately, current surveys of the distant universe continue to
provide us with better constraints on the volume density of galaxies
over a wide range in luminosity and redshift \citep[e.g.,
][]{PO14,RB15}.  Both very sensitive observations with the Hubble
Space Telescope and wide field observations from ground-based
instruments are key to obtaining these improved constraints on the
volume densities.  Quantifying the volume density of ultra-faint
galaxies is particularly important for determining the impact of
galaxies to reionizing the universe.

\section{Galaxies as a Potential Source of the Cosmic Ionizing Emissivity}
\label{sec:2}

Obtaining direct constraints on the ionizing UV radiation from
galaxies escaping into the intergalactic medium is quite difficult.
Direct detection of this radiation can barely be done for galaxies at
$z\sim1$-3 or even in the nearby universe.  For galaxies at $z\geq6$
where reionization occurs, this endeavor is exceedingly challenging,
as any ionizing radiation must first redshift through the dense
Lyman-series forest in the high-redshift universe prior to any
attempted observation.  Indirect methods, such as use of the proximity
effect \cite{RC82,SB88}, to constrain the ionizing radiation from
galaxies may hold some promise, but that technique has largely been
used in the context of such radiation from quasars.

Because of the difficulties in setting direct constraints on the total
ionizing radiation coming from galaxies, astronomers try to estimate
this total as the product of three quantities, the rest-frame $UV$
luminosity density $\rho_{UV}$, an efficiency factor in converting the
$UV$ luminosity to Lyman-continuum emission $\xi_{ion}$, and the
escape fraction $f_{esc}$.  Existing observations allow us to place
firm lower limits on the first of these quantities, the $UV$
luminosity density $\rho_{UV}$, while the third quantity here, the
escape fraction $f_{esc}$, is much more difficult to accurately
constrain.  The escape fraction can be defined in a variety of ways
but roughly expresses the relative fraction of
Lyman-continuum-ionizing photons escaping from galaxies to the
fraction of $UV$-continuum photons which escape.

\subsection{UV-continuum Luminosity Density}
\label{sec:21}

Of those quantities relevant to galaxy's role in reionizing the
universe, the most straightforward quantity to constrain is the
rest-frame $UV$ luminosity density $\rho_{UV}$.  This density
quantifies the total luminosity of galaxies at $UV$-continuum
wavelengths in a given comoving volume of the universe.

To quantify the luminosity density $\rho_{UV}$ at a given epoch,
astronomers take the volume density of galaxies they derive from
searches for galaxies as a function of luminosity (i.e., the $UV$
luminosity function), multiply this volume density by the luminosity,
and then integrate this product over the full range of observed (and
expected) galaxy luminosities:
\begin{equation}
\rho_{UV} = \int _{L_{min}} ^{L_{max}} \,\phi(L) LdL
\label{eq:intl}
\end{equation}
where $L$ is the $UV$ luminosity of galaxies, $\phi(L)$ is the volume
density of galaxies as a function of luminosity, and $L_{min}$ and
$L_{max}$ are the lowest and highest luminosities that galaxies can
attain.

As is apparent from the equation above, deriving the $UV$ luminosity
density $\rho_{UV}$ at a given epoch is nominally a very rudimentary
calculation to perform, after one derives the $UV$ luminosity function
from one or more observational probes.  In practice, however, this
endeavor is not entirely straightforward, which is a direct
consequence of the apparently large population of galaxies with
luminosities fainter than what we can readily probe with existing data
sets.  Observationally, there is no credible evidence for a possible
cut-off in the $UV$ luminosity function towards the faint end of what
can be currently observed.  Additionally, from simple theoretical
models, one could reasonably expect galaxies to efficiently form
$\sim$50 to 1000$\times$ fainter than the current observational
limits.

\begin{figure}[h]
\hspace{2cm}\includegraphics[scale=0.4]{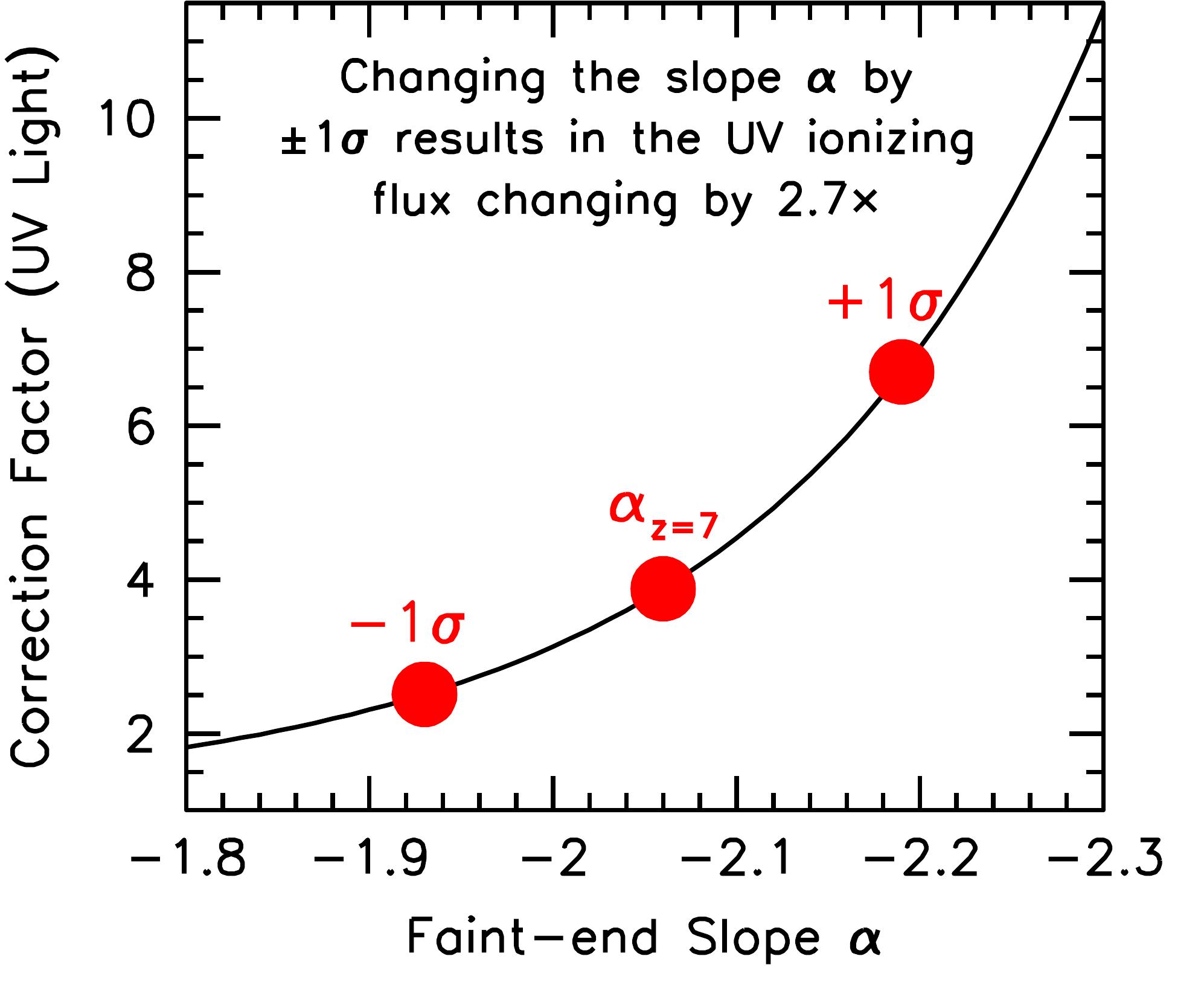}
\caption{The dependence of the integrated UV ionizing flux on the
  faint-end slope $\alpha$ (\S\ref{sec:23}).  The correction factors
  needed to convert the observed $UV$ photon density to the total
  density for a given faint-end slope $\alpha$ (integrating to the
  expected theoretical cut-off in the LF at $-$10$\,$mag) are shown
  for the current best estimate of the slope $\alpha$ at $z\sim7$ and
  for the upper and lower $\pm$1$\sigma$ limits of $\alpha\sim-1.93$
  and $\alpha\sim-2.19$ \citep{RB15}.  The current uncertainties in
  the correction factor are large (the ratio between the upper and
  lower 1$\sigma$ correction factors is a factor of $\sim$3).}
\label{fig:totld}
\end{figure}

The importance of faint galaxies for driving the reionization of the
universe lies in their large volume densities.  Extrapolating the
observed luminosity function to lower luminosities (e.g., $-$10 mag:
$\sim$10$^4$ fainter than $L^*$ galaxies) suggest that these
ultra-faint (and largely individually undetectable) galaxies could
produce $\sim$7$\times$ as much light as the galaxies that we can
probe directly with currently observable surveys \citep[e.g.,
][]{RB12A} (Figure~\ref{fig:totld}).

\subsection{Evolution of the $UV$ LF from Early Times}
\label{sec:22}

One particularly important aspect of ascertaining whether galaxies can
be successful in reionizing the universe involves a characterization
of their evolution with cosmic time.  The rate of evolution can have a
big impact on how luminous galaxies are in the early universe and this
will affect how many ionizing photons they produce.

After more than 10 years of careful quantitative analyses of large
samples of $z\sim4$-8 galaxies, the evolution of the $UV$ LF is now
very well characterized over the redshift range $z\sim4$ to $z\sim8$.
Essential for these analyses are the very sensitive, wide-area images
of the distant universe in multiple wavelength channels, stretching
from near-UV wavelengths into the infrared.

Due to the unique colors and spectral shape of young star-forming
galaxies in the $z\geq4$ universe, one can take advantage of the rich
multi-wavelength information in deep imaging observations to identify
large numbers of largely robust $z\sim4$-8 galaxies.  The general
technique used to identify distant galaxies is called the Lyman-break
technique and has been demonstrated to work very efficiently from
important pioneering work in the mid-1990s on galaxies at $z\sim3$
\citep{CS96,CS03}, with subsequent demonstrations on thousands of
galaxies at $z=3$-6 \citep[e.g., ][]{EV08,DS10} to the current
high-redshift record-holder $z=7.73$ \citep{PO15}.  Application of
techniques like the Lyman-break technique and more generally
selections using photometric-redshift estimators to data both from
large telescopes on the ground and from the Hubble Space Telescope
\citep[e.g., ][]{RV10,RB15} have made it possible to construct
samples with extremely large numbers ($>$10$^4$) of galaxies.

\begin{figure}[h]
\includegraphics[scale=0.6]{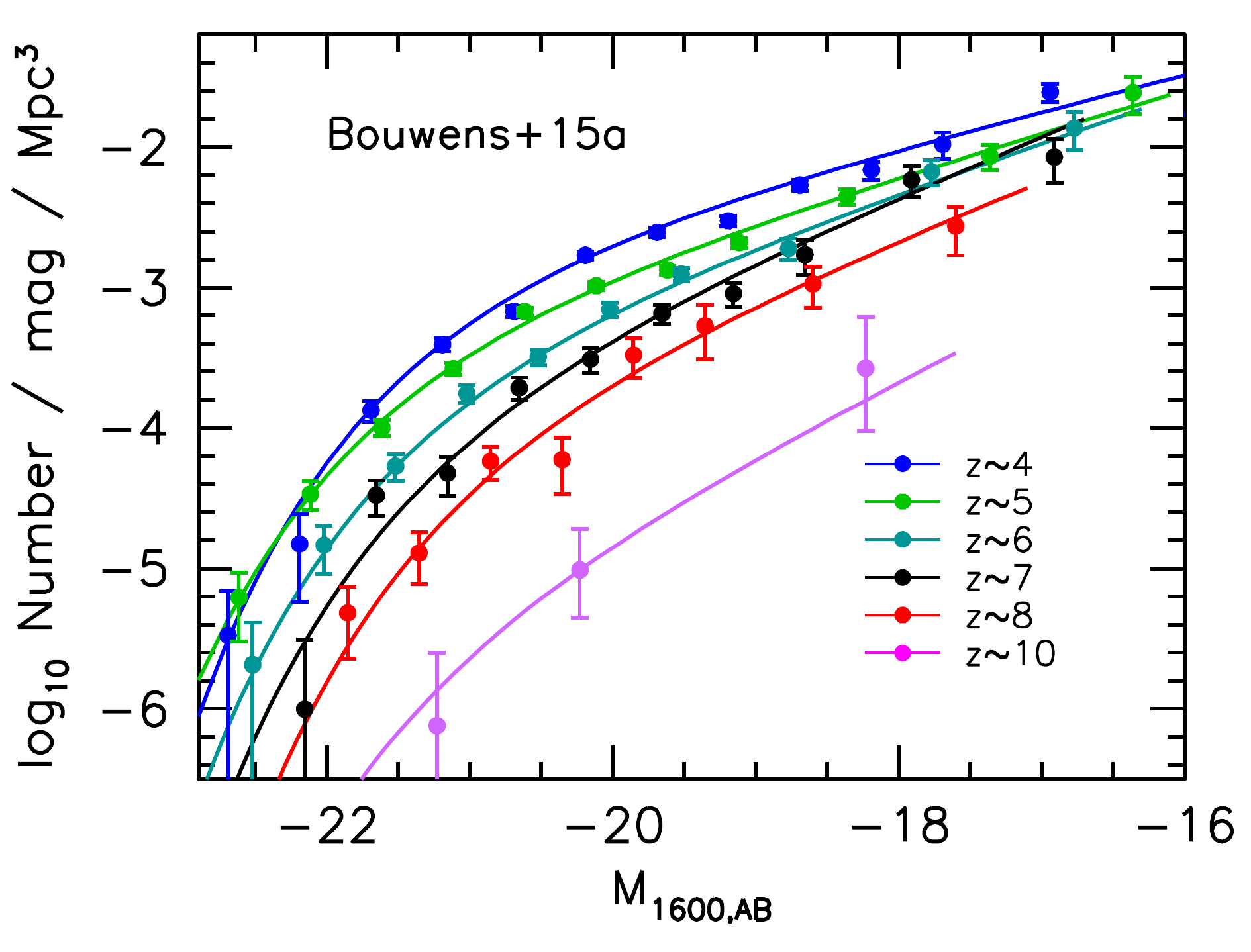}
\caption{Current state-of-the-art determinations of the $UV$
  luminosity functions at $z\sim4$ (\textit{blue}), $z\sim5$
  (\textit{green}), $z\sim6$ (\textit{cyan}), $z\sim7$ (\textit{red}),
  $z\sim8$ (\textit{black}), and $z\sim10$ (\textit{magenta}) using
  the full data sets over the CANDELS + HUDF + HUDF-parallel +
  BoRG/HIPPIES fields \citep{RB15} (\S\ref{sec:22}).  The solid
  circles represent stepwise maximum-likelihood determinations of the
  LFs while the solid lines are the Schechter function determinations.
  These luminosity functions allow for an accurate quantification of
  the luminosity density of galaxies in the rest-frame $UV$ from $z=4$
  to $z=10$ and therefore also an estimate of ionizing photons
  available to reionize the universe.}
\label{fig:lfall10}       
\end{figure}

The most valuable observations for constraining the luminosity
functions are those that probe the volume density of the faintest
sources and those that probe the volume density of the brightest
sources.  Constraints on the volume density of the faintest sources
have typically come from the optical and near-infrared observations of
the Hubble Ultra Deep Field (HUDF) with the Hubble Space Telescope
with the Advanced Camera for Surveys Wide-Field Camera and the
near-infrared channel on the Wide Field Camera 3.  Most of the
observations were obtained as part of the original HUDF campaign in
2004 \citep{SB06} and as part of the HUDF09 and HUDF12 campaigns in
2009-2012 \citep{RB11,RE13}.  These observations are effective in
detecting soruces to 30 AB mag, equivalent to a luminosity of $-16$
mag at $z\sim4$ and $-17$ mag at $z\sim7$ \citep{RB07,MS13,RM13,RB15}.
Figure~\ref{fig:lfall10} shows one recent state-of-the-art compilation
of LF determinations.

Constraints on the volume density of bright galaxies have
predominantly come from two sources: (1) the wide-area ($\sim$900
arcmin$^2$) optical and near-infrared observations from the Great
Observatories Origins Deep Survey \citep[GOODS: ][]{MG04} and CANDELS
program \citep{NG11,AK11} and (2) square-degree ground-based search
fields like the UKIDSS Ultra Deep Survey, UltraVISTA \citep{HM12}, and
the Canada-France-Hawaii Telescope deep legacy survey fields
\citep{RV10,CW13}.  These wide-area programs generally have been
successful in finding a few very bright star-forming galaxies to
$\sim$23-24 mag.  The search results have important implications for
the build-up of bright galaxies and the evolution of the
characteristic luminosity $L^*$, but do not have an especially
meaningful impact on the overall background of ionizing photons in the
$UV$.

Very good agreement is found between independent determinations of the
rest-frame $UV$ luminosity functions of galaxies at $z\sim4$-10
\citep{RB15,SF15,RBow14,RBow15}.  Most of the debate in the literature has
revolved about how the evolution can be best represented in terms of
the various Schechter parameters (e.g., \cite{RB12A,RM13}
vs. \cite{RB15}).\footnote{This debate has continued for many years
  due to the significant degeneracies that exist between the different
  Schechter parameters (and also partially due to questions about
  whether the functional form of the LF is Schechter at $z\geq7$:
  e.g., \cite{RBow14,RB15}).}  Agreement between different
determinations is generally good, when expressed in terms of the
implied UV luminosity densities.

The $UV$ LF appears to evolve with redshift in a reasonably smooth
manner, with the characteristic magnitude $M^*$, the logarithm of the
normalization $\phi^*$, the faint-end slope $\alpha$, and logarithm of
the luminosity density all varying with redshift in an approximately
linear manner from $z\sim8$ to $z\sim4$ \citep{RB08,RB11}.

\begin{figure}[h]
\hspace{1.3cm}\includegraphics[scale=0.6]{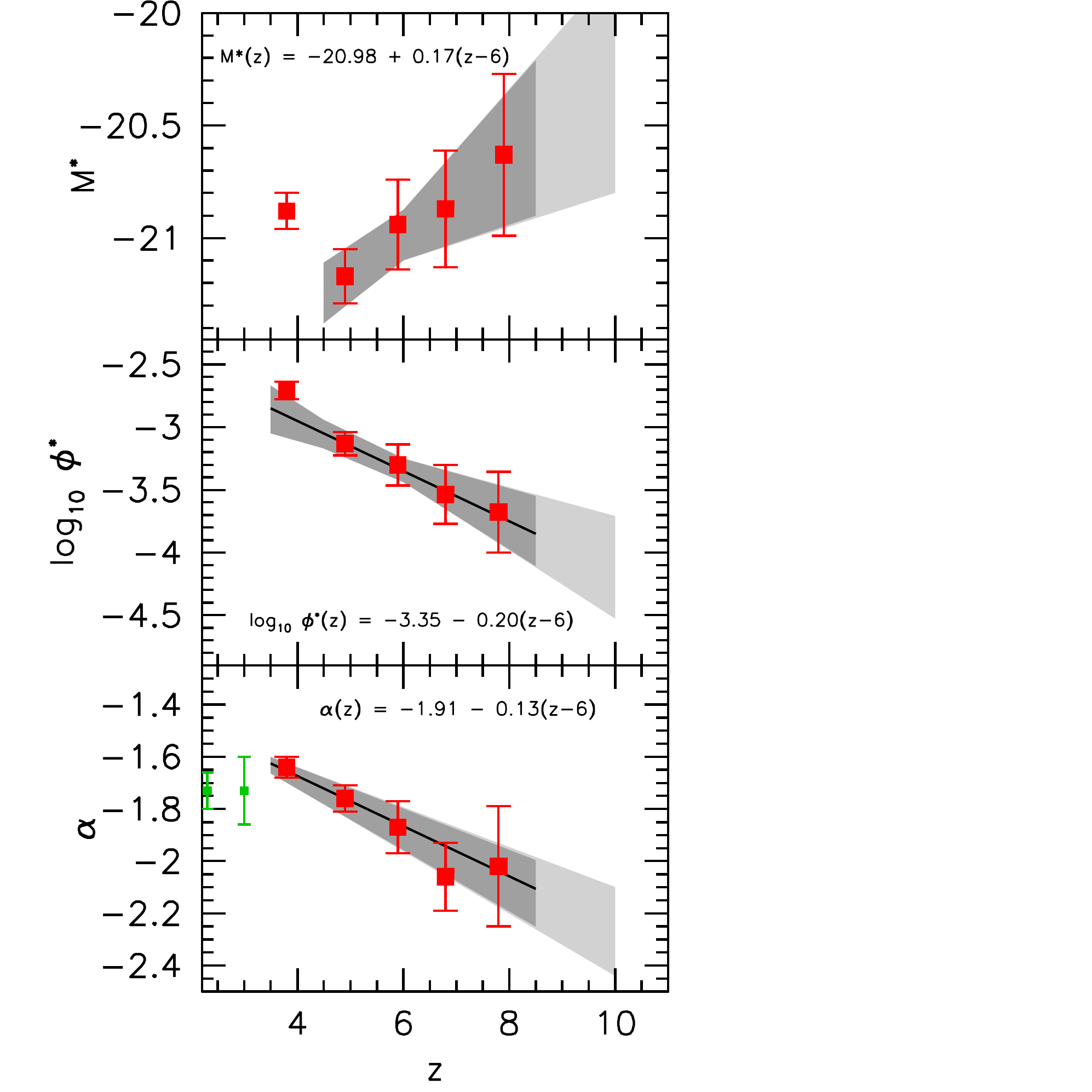}
\caption{Best-fit constraints on the evolution of the three Schechter
  parameters $M^*$ (upper), $\phi^*$ (middle), and $\alpha$ (lower
  panels) to $z>5$ using recent state-of-the-art luminosity function
  determinations from \cite{RB15} (\S\ref{sec:22}).  The solid line is
  a fit of the $z\sim4$-8 faint-end slope determinations to a line,
  with the 1$\sigma$ errors (gray area: calculated by marginalizing
  over the likelihoods for all slopes and intercepts).  Also shown are
  the faint-end slope determinations from \cite{NR09} at $z\sim2$-3
  (\textit{green squares}).  The evolutionary trend most relevant for
  galaxies' reionizing the universe are the changes in the faint-end
  slope $\alpha$.  The best-fit trend with redshift (from $z\sim5$ to
  $z\sim8$) is $d\alpha/dz=-0.13\pm0.03$.  Strong evidence
  ($3.4\sigma$) is found for a steepening of the $UV$ LF from $z\sim8$
  to $z\sim4$.}
\label{fig:schevol}
\end{figure}

Here we provide one such fit to the evolution of the $UV$ LF with
redshift:
\begin{eqnarray*}
M_{UV} ^{*} =& (-20.99\pm0.10) + (0.18\pm0.06) (z - 6)\\
\phi^* =& (0.44_{-0.10}^{+0.11}) 10^{(-0.20\pm0.05)(z-6)}10^{-3} \textrm{Mpc}^{-3}\\
\alpha =& (-1.91\pm0.05) + (-0.13\pm0.03)(z-6)
\label{eq:empfit}
\end{eqnarray*}
The above fit makes full use of the likelihood contours derived by
\cite{RB15} for the galaxy LFs at $z\sim5$-8 and is illustrated in
Figure~\ref{fig:schevol}.  We have elected to make exclusive use of
the constraints from \cite{RB15} at $z=5$, $z=6$, $z=7$, and $z=8$ due
to the relatively smooth increase in the characteristic luminosity
$L^*$ and normalization $\phi^*$ and flattening of the faint-end slope
$\alpha$ observed over this redshift range (see also \cite{RBow15,SF15}).
These trends appear to be significant at $\geq3\sigma$ in all three
cases.

\subsection{Faint-end Slope and Its Evolution}
\label{sec:23}

In deriving estimates of the total luminosity density $\rho_{UV}$ in
the largely unobservable population of ultra-faint galaxies, there are
two considerations which are important.  The first of these
considerations is the faint-end slope $\alpha$ of the luminosity
function -- which gives the power-law relationship between the volume
density of galaxies at some epoch and the $UV$ luminosity of galaxies
at that epoch.  The second of these considerations is the minimum
luminosity at which we would expect galaxies to efficiently form.  We
will discuss the first of these considerations in this subsection and
discuss the second in \S\ref{sec:252}.

To illustrate the importance of the faint-end slope $\alpha$ for
estimates of the total luminosity density, we present in
Figure~\ref{fig:totld} how the ratio of the total-to-observed
luminosity density depends on the faint-end slope assumed.  Even small
uncertainties in the faint-end slope $\alpha$ can have a big impact on
the inferred total luminosity density, as the faint-end slope $\alpha$
approaches a value of $\sim-2$ where the integral for the luminosity
density (Eq.~\ref{eq:intl}) formally becomes divergent.

\begin{figure}[h]
\hspace{1.5cm}\includegraphics[scale=0.65]{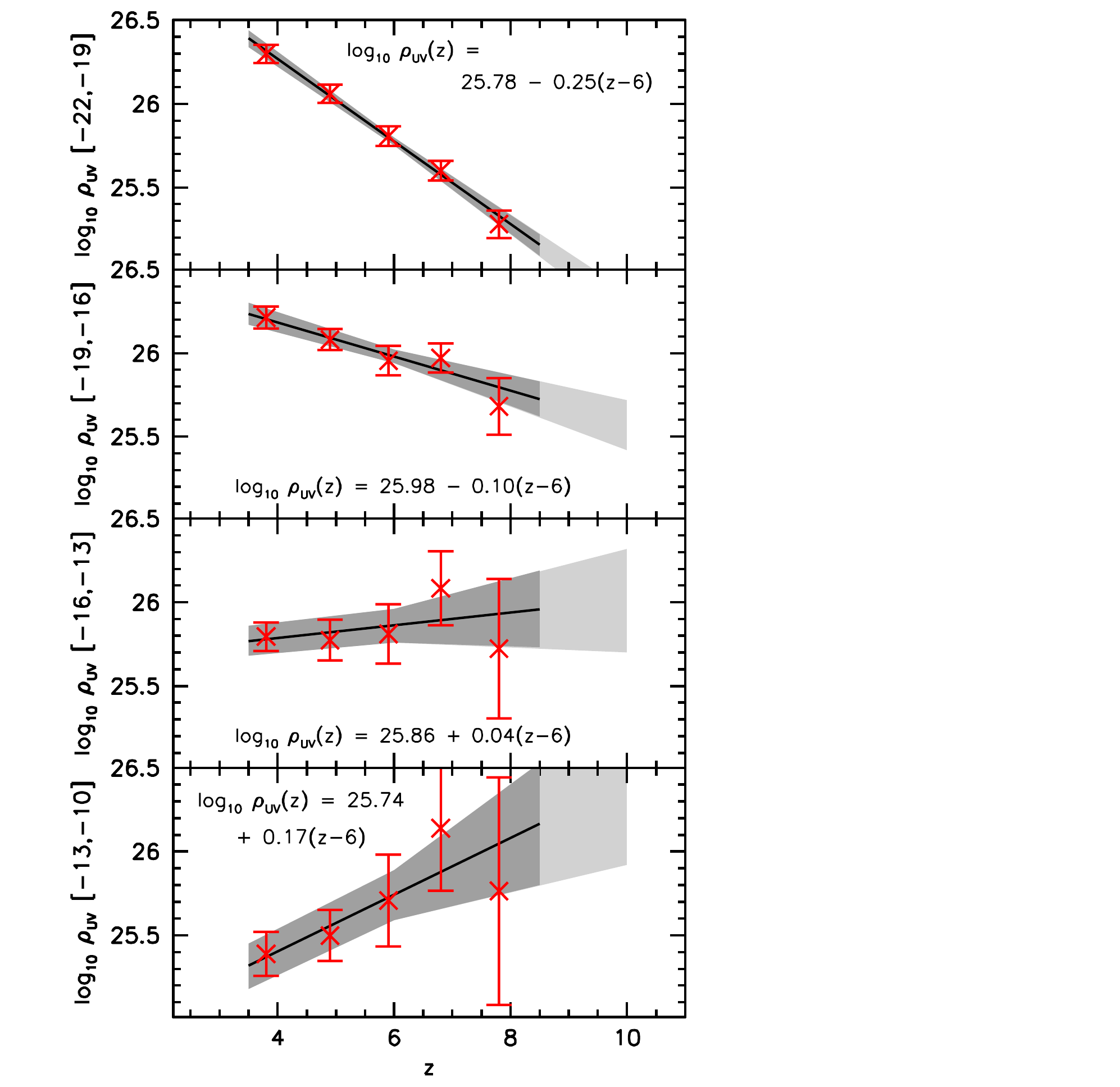}
\caption{The evolution of the $UV$ luminosity density $\rho_{UV}$ over
  various intervals in absolute magnitude, i.e., $-22<M_{UV,AB}<-19$
  (uppermost panel), $-19<M_{UV}<-16$ (second uppermost panel),
  $-16<M_{UV}<-13$ (second lowest panel), $-13<M_{UV}<-10$ (lowest
  panel), based on the recent comprehensive determinations of the
  $z=4$-8 LFs from \cite{RB15} (\S\ref{sec:26}).  The red crosses and
  error bars give the determinations based on the LFs at specific
  redshifts, while the grey-shaded region gives the 68\% confidence
  intervals assuming a linear dependence of $\log_{10} \rho_{UV}$ on
  redshift.  Constraints on the evolution of this luminosity density
  over the magnitude range $-22<M_{UV}<-16$ are largely based on the
  direct search results from Bouwens et al.\ \citep{RB15} while the
  results over the magnitude range $-16<M_{UV}<-10$ are based on
  extrapolations of the LF beyond what can be directly detected.  If
  galaxy formation is particularly inefficient faintward of some
  luminosity, then we would expect that the luminosity densities
  derived here to be an overestimate.}
\label{fig:ldevol} 
\end{figure}

The first accurate measurement of the faint-end slope $\alpha$ for
galaxies in the high-redshift universe were performed by \cite{CS99}
at $z\sim3$ in 1999.  A relatively steep slope of $\sim-1.6$ was
found, which implied that the surface density of galaxies on the sky
approximately doubled for every $\Delta m = 1$ increase in apparent
magnitude.  Such steep faint-end slopes imply that lower luminosity
galaxies contribute significantly to the overall luminosity density of
the universe.

Precise measurements of the faint-end slope at other redshifts
($z\sim1$-5) took somewhat longer to become a reality.  Most studies
found faint-end slopes ranging from $\sim-$1.5 to $\sim-$1.7
\citep{SA05,RB07,NR09,PO10}.  At higher redshifts, determinations of
the faint-end slope were much less certain, but there was general
agreement that the faint-end slope $\alpha$ was at least as steep at
$-1.7$ \citep{HY04,RB07}.

The first real progress in deriving the faint-end slope at $z\sim 7$-8
came with the availability of deep WFC3/IR data over the HUDF
\citep{RB11}.  Combining constraints on the volume density of luminous
$z\sim7$-8 galaxies, with constraints on the volume density of
significantly fainter galaxies from the new 192-orbit HUDF09 data set,
\cite{RB11} presented the first tantalizing evidence for a further
steepening of the LF at $z>6$.  A faint-end slope $\alpha$ of
$-2.01\pm0.21$ was found at $z\sim7$ and $-1.91\pm0.34$ at $z\sim8$.

Towards the end of 2012, even deeper observations were obtained over
the HUDF as a result of the HUDF12 campaign \citep{RE13}, extending
the depth in the $1.1\mu$m $Y_{105}$-band data by $\sim$0.7 mag and
adding observations at $1.4\mu$m in the $JH_{140}$ band.  These new
observations strengthened existing evidence at $z\sim7$ and $z\sim8$
that the faint-end slope at $z\sim7$-8 was steep \citep{RM13,MS13}

The first definitive clarification of the evolution came with the $UV$
LF determinations from \cite{RB15}.  In that study, LFs were derived
from large $z\sim4$, 5, 6, 7, 8, and 10 samples identified from the
$\sim$1000 arcmin$^2$ CANDELS + HUDF + HUDF-parallel + BoRG/HIPPIES
data set \citep{MT11,HY11} and possible evolution in Schechter
parameters considered.  The faint-end slopes derived at high
redshifts, i.e., $z\sim7$ and $z\sim8$, were much steeper
($\Delta\alpha\sim0.4$) than those found at the low end of the
redshift range considered by \cite{RB15} (3.4$\sigma$ significance).
This is illustrated in the lowest panel of Figure~\ref{fig:schevol}.
The advantage of mapping out the evolution of the LF over such a wide
range in redshift using the same techniques and observational data is
that it largely guarantees the results will be free of systematic
errors.

The much steeper faint-end slopes to the $UV$ LFs at early times has
one particularly important implication.  Lower-luminosity galaxies
will evolve much less in their overall volume density with cosmic time
than more luminous galaxies, and consequently still be very abundant
in the early universe (Figure~\ref{fig:ldevol}).  As a result, such
galaxies are expected to contribute the vast majority of the photons
that reionize the universe.

\begin{figure}[h]
\hspace{1.5cm}\includegraphics[scale=0.6]{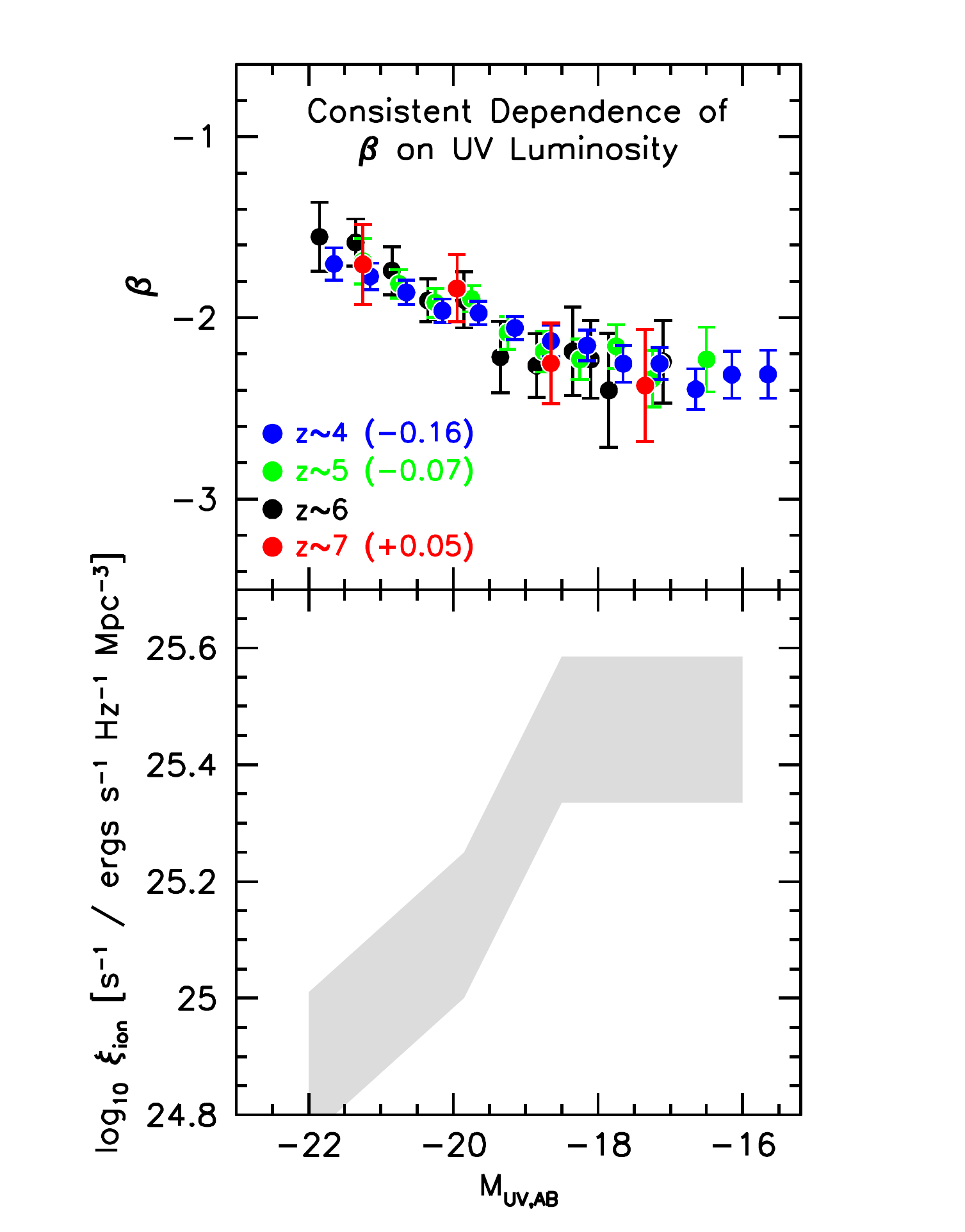}
\caption{(\textit{upper}) Mean $UV$-continuum slope $\beta$ of
  star-forming galaxies at $z\sim4$ (\textit{blue}), $z\sim5$
  (\textit{green}), $z\sim6$ (\textit{black}), and $z\sim7$
  (\textit{red}) versus their measured $UV$ luminosities, as derived
  by \cite{RB14} (\S\ref{sec:24}).  The mean $\beta$'s presented here
  are offset slightly depending on the redshift to illustrate the
  similarity of $\beta$ vs. $M_{UV}$ relations for galaxies in all
  four redshift intervals.  The $UV$-continuum slope $\beta$ exhibits
  a much stronger dependence on luminosity for galaxies brightward of
  $-19$ mag than it does faintward of this luminosity.  A number of
  other recent studies \citep{RB12B,AR14} have recovered a very
  similar dependence of $\beta$ on $UV$ luminosity, as shown here.
  (\textit{lower}) Suggested conversion factors $\xi_{ion}$ to use in
  transforming the observed luminosity density in the $UV$ continuum
  to the equivalent density in ionizing photons.  The preferred
  conversion factor can be estimated based on the mean $UV$-continuum
  slopes $\beta$ derived for Lyman-break galaxies at a given redshift
  and luminosity (see Figure~\ref{fig:xi}).  The conversion factors
  suggested here are derived from mean $\beta$'s presented in the
  upper panel.}
\label{fig:beta47}
\end{figure}

\subsection{Estimating the Production Rate of Lyman-Continuum Photons}
\label{sec:24}

To determine if the observed galaxy population can be successful in
reionizing the universe, we must convert the overall luminosity
density $\rho_{UV}$ in the $UV$-continuum ($\sim$1600 \AA) to the
equivalent luminosity density in Lyman-continuum photons ($\leq$912
\AA).  As this involves an extrapolation of the observed $UV$ light to
bluer wavelengths, one might expect the observed $UV$ colors to
provide us with the necessary information to perform this
extrapolation more accurately.

It is conventional to model the spectrum of galaxies in the rest-frame
$UV$ as a power-law such that $f_{\lambda} \propto \lambda^{\beta}$
(or equivalently $f_{\nu} \propto \nu^{-(\beta+2)}$) where $\beta$ is
the $UV$-continuum slope and $\lambda$ is the wavelength.  While the
spectrum of star-forming galaxies in the $UV$ continuum cannot be
perfectly described using a power-law parameterization, such a
parameterization generally works for most of the spectral range to
within $\pm$20\%.

Over the last few years, significant effort has been devoted to
quantifying the $UV$-continuum slope $\beta$ distribution of galaxies
as a function of both the $UV$ luminosity and redshift of galaxies
\citep{GM99,KA00,ES05,RB09,RB12B,RB14,SW11,MC12,JD12,SF12A,AR14}.

In general, the $UV$-continuum slopes $\beta$ of galaxies at
$z\sim4$-8 have been found to have a mean $\beta$ of $\sim-$1.6 at
high luminosities and slowly trend towards bluer $\beta$'s of
$\sim-$2.2 at the lowest luminosities \citep{RB12B,RB14,AR14}.  The
relationship is remarkably similar for galaxies at $z\sim4$, $z\sim5$,
$z\sim6$, and $z\sim7$, as illustrated in Figure~\ref{fig:beta47}.

\begin{figure}[h]
\hspace{1.5cm}\includegraphics[scale=0.6]{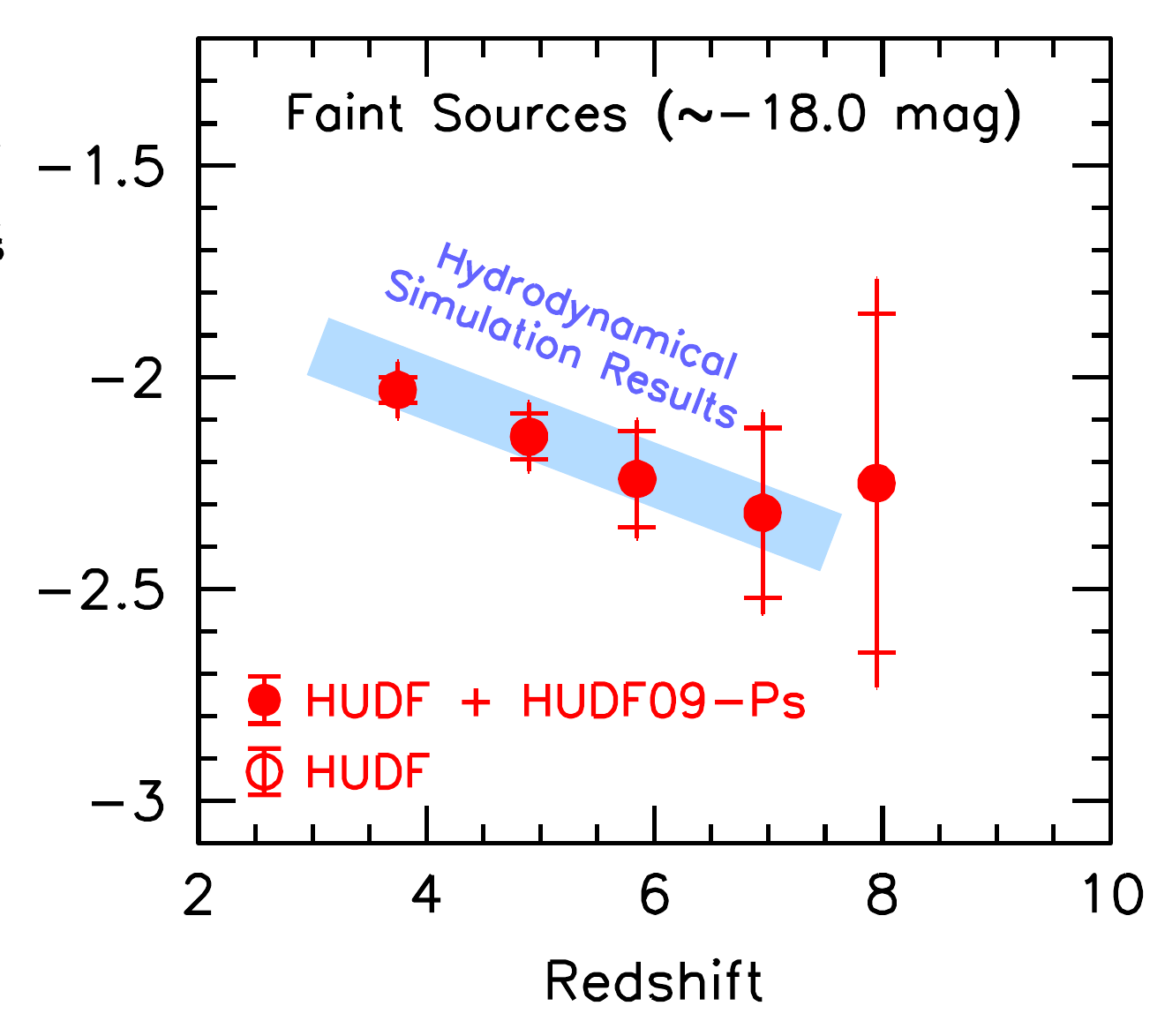}
\caption{The biweight mean $UV$-continuum slopes $\beta$ observed by
  \cite{RB14} for faint ($-19\leq M_{UV,AB}\leq-17$) $z\sim4$,
  $z\sim5$, $z\sim6$, $z\sim7$, and $z\sim8$ galaxies in the HUDF and
  HUDF-parallel fields (\textit{red solid circles}: \S\ref{sec:24}).
  $1\sigma$ uncertainties on each of the determinations are also
  shown, for the statistical uncertainties alone (\textit{including
    hashes at the ends}) and including the systematic uncertainties
  (\textit{not including hashes}).  The expectations from
  hydrodynamical simulations \citep{KF11} are shown for comparison
  with the thick light-blue line.  The weak evolution observed in
  $\beta$ is expected to result in $\sim$50$\pm$25\% more ionizing
  radiation for faint galaxies at $z\sim7$-8 than at $z\sim4$.}
\label{fig:beta_evol}
\end{figure}

There is some evidence for a weak evolution
($d\beta/dz\sim-0.10\pm0.05$) in $\beta$ with redshift for lower
luminosity galaxies, from $\beta\sim-2.3$ for $z\sim7$-8 galaxies to
$\beta\sim-2.1$ for $z\sim4$ galaxies
\citep{RB12B,RB14,SF12A,NH13,PK14} (see Figure~\ref{fig:beta47} and
\ref{fig:beta_evol}).  The observed evolution is consistent with that
expected for the zero-attenuation $UV$ slopes from simulations
\citep{SW13} (see also \cite{KF11}).  The scatter in the
$UV$-continuum slopes is $\sim$0.35 for the most luminous galaxies
\citep{RB09,RB12B,MC12}, but appears to decrease to $\sim$0.15 for the
lowest-luminosity galaxies \citep{AR14}.

\begin{figure}[h]
\hspace{1.5cm}\includegraphics[scale=0.76]{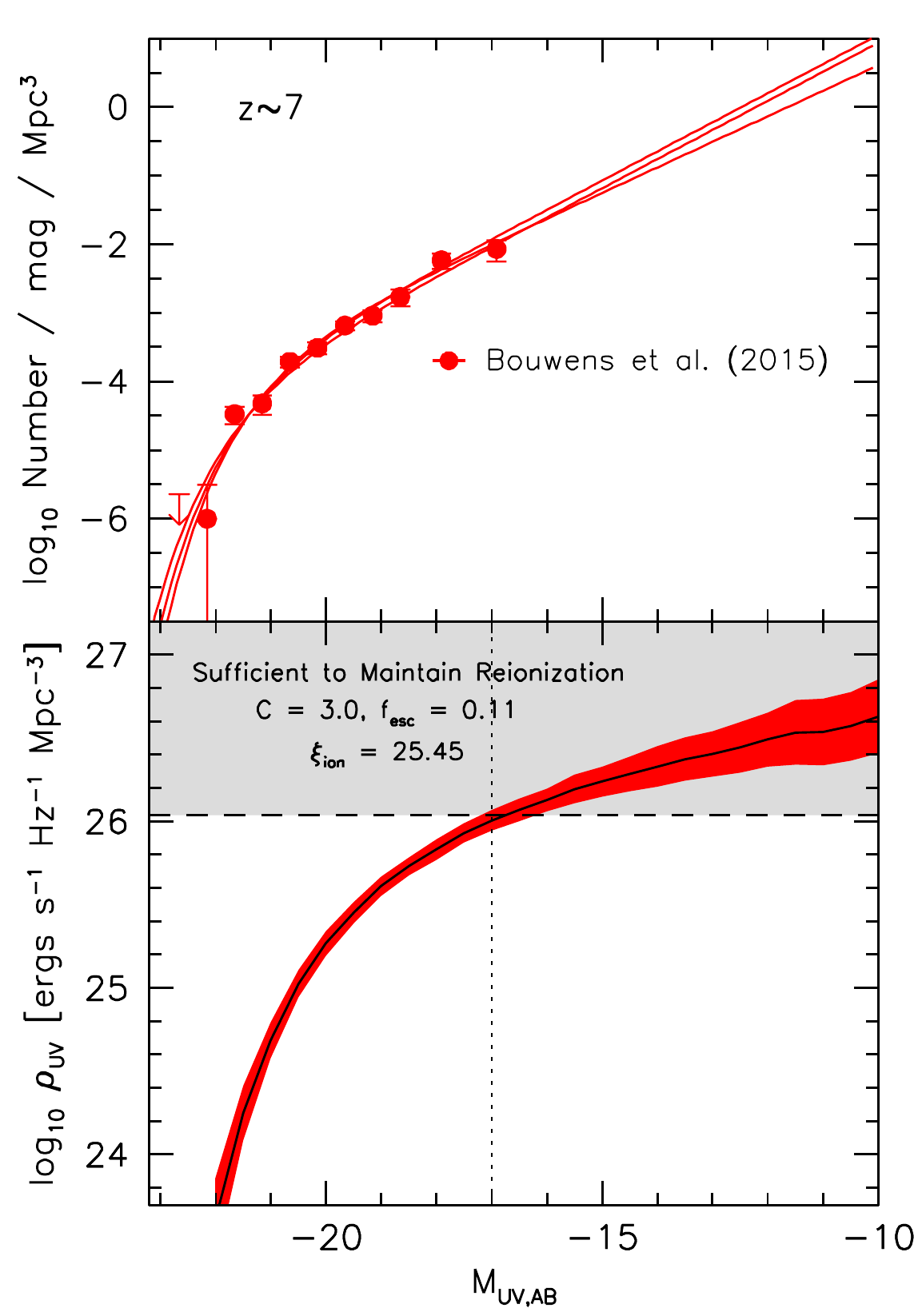}
\caption{\textit{(top)} Best-fit constraints on the galaxy luminosity
  function at $z\sim7$ in the rest-frame ultraviolet from \cite{RB15}
  using observations from the full CANDELS, HUDF, and HUDF parallel
  programs (\S\ref{sec:22}).  Shown are both binned and parameterized
  constraints on the luminosity function, with uncertainties in the
  extrapolated relation represented by the different lines.
  \textit{(bottom)} Maximum likelihood constraints (\textit{black
    line}) on the $UV$ luminosity density $\rho_{UV}$ integrated to
  different lower luminosity limits presented relative to the
  requisite luminosity density in the $UV$ needed to reionize the
  universe.  The luminosity densities preferred at 68\% confidence
  \citep{RB15} are indicated with the red-shaded
  regions.\label{fig:comp7}}
\end{figure}

\begin{figure}[h]
\hspace{1cm}\includegraphics[scale=0.5]{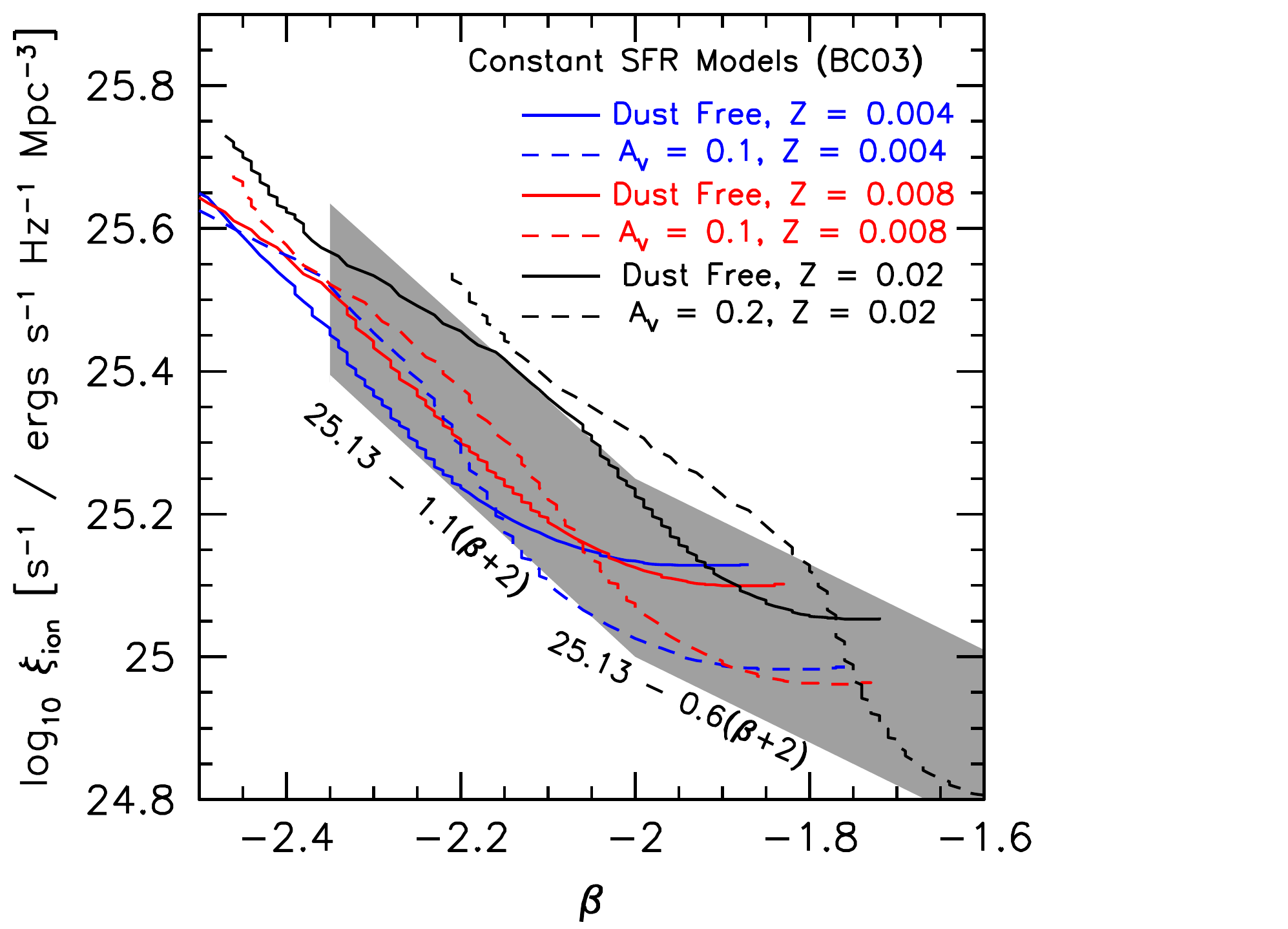}
\caption{A determination of how the production efficiency $\xi_{ion}$
  of Lyman-continuum photons per unit $UV$ luminosity at 1600\AA$\,\,$
  depends on the $UV$-continuum slope $\beta$ (\S\ref{sec:24}).  These
  efficiencies are calculated from the Bruzual \& Charlot \citep{GB03}
  spectral synthesis library assuming a constant star formation rate
  adopting three different metallicities (0.004$Z_{\odot}$,
  0.008$Z_{\odot}$, and 0.02$Z_{\odot}$) and a wide range in ages
  (ranging from an essentially instantaneous burst [10$^5$ years] to a
  stellar population of similar age to the universe itself [10$^{10}$
    years]).  Both the case of no dust content and $\tau_{V} =
  0.1/0.2$ \citep{SC00} is considered, as indicated on the figure.
  $\beta$ is computed over the spectral range 1700\AA$\,\,$to 2200\AA.
  The shaded envelopes indicate the approximate dependence of
  $\xi_{ion}$ on $\beta$.}
\label{fig:xi}
\end{figure}

Since essentially all of the $UV$ light produced by galaxies in the
$z\sim6$-8 universe derives from galaxies at very low luminosities
(Figure~\ref{fig:comp7}), it is reasonable to use the $\beta$'s
measured for lower luminosity galaxies to convert the luminosity
densities in the $UV$ continuum into the equivalent density of
ionizing photons.  The conversion factor can be estimated using the
spectral synthesis models of \cite{GB03}, assuming a constant
star-formation history and a variety of different ages and
metallicities for the stars, as well as a modest amount of dust
content.  The conversion factors and $\beta$'s computed for many
different model spectra are presented in Figure~\ref{fig:xi}.
Estimates of these conversion factors were previously estimated by
\cite{BR13} using earlier measurements of $\beta$ for $z\sim7$
galaxies \citep{JD13}.

Significantly enough, for the faint population of star-forming
galaxies at $z\sim5$-8, the mean $UV$-continuum slope $\beta$ imply
that the conversion factor to the ionizing luminosity density
$\xi_{ion}$ is approximately equal to $10^{25.45}$ s$^{-1}$ $/$
(ergs\,s$^{-1}$ Hz$^{-1}$).  This is somewhat higher than the
$10^{25.3}$ s$^{-1}$ $/$ (ergs\,s$^{-1}$ Hz$^{-1}$) conversion factor
adopted by \cite{RB12A} and \cite{MK12} and the $10^{25.2}$ s$^{-1}$
$/$ (ergs\,s$^{-1}$ Hz$^{-1}$) conversion factor adopted by
\cite{PM99} and \cite{BR13}.

\subsection{Key Uncertainties in Computing the Ionizing Photon Density Contributed by Galaxies}
\label{sec:25}

\subsubsection{Lyman-Continuum Escape Fraction}
\label{sec:251}

For $z>6$ galaxies to have been successful in reionizing the universe,
a modest fraction of the ionizing radiation emitted from their hot
stars, i.e., $\geq$10\%, must escape into the intergalactic medium.
This general expectation has significantly motivated the search for
such escaping radiation both at intermediate redshifts and in the
nearby universe.  Direct searches for such radiation from $z>6$
galaxies, themselves, would also be interesting and provide the most
relevant information on this issue, but are not really feasible, in
that any escaping radiation from $z>6$ galaxies would need to
successfully redshift through the thick Lyman-series forest to allow
for detection.

Remarkably enough, searches for leaking ionizing radiation from
galaxies in the nearby universe has been largely a frustrating
activity, with little success.  Most such studies only result in ever
more stringent upper limits on the ionizing radiation escaping from
galaxies \citep{CL95,JD01,EG02,MM03,BS07,JG07,LC09,BS10}.  Perhaps the
most promising results in searches for escaping radiation from
galaxies have come from the very compact star-forming galaxies, with a
particularly dominant central component, where galaxies show evidence
for having a low covering fraction of neutral hydrogen gas
\citep{TH11} and where one galaxy shows direct evidence for $\sim$20\%
of the ionizing radiation escaping from the galaxy.

Efforts to identify significant leaks of ionizing radiation from
$z\sim2$-3 galaxies have been seemingly more successful than at
$z\sim0$-1.  Though there is significant debate in the literature
regarding the precise value of the escape fraction $f_{esc}$, the
reported values for the escape fraction range from to values of 2-5\%
\citep{EV12} to 10-30\% \citep{DN13,RM13,JC14,BS15}.

The substantial scatter in estimated values for the escape fraction is
a direct consequence of the challenges inherent in accurately
quantifying it.  A precise determination of this fraction require
accurate redshift measurements for a representative sample of
intermediate-redshift sources, as well as a measurement of the flux
blueward of the Lyman break.  One issue of particular importance is
that of foreground sources at lower redshift lying almost directly in
front of star-forming galaxies and therefore effectively mimicking the
signature of Lyman-continuum emission \citep[e.g., ][]{DN11,EV12}.
Another issue regards the possible exclusion of those sources showing
the strongest Lyman-continuum emission from high-redshift samples
\citep{JC14}.  Both issues continue to be the subject of debate in the
literature \citep{BS15}.

Some discussion about how the escape fraction is defined is required,
as it can described in a variety of ways in the literature.
Throughout most of the present chapter and in most self-consistent
reionization models of the universe, the escape fraction $f_{esc}$
discussed is the so-called ``relative escape fraction'' $f_{esc,rel}$
\citep{CS01,BS10}:
\begin{equation}
f_{esc,rel} = \frac{f_{esc}^{LyC}}{f_{esc}^{UV}} =
\frac{(L_{UV}/L_{LyC})_{intr}}{(F_{UV}/F_{LyC})_{corr}}
\end{equation}
where $f_{esc}^{LyC}$ and $f_{esc}^{UV}$ describes the fraction of
Lyman Continuum and $UV$ continuum photons, respectively, emitted from
stars that successfully escape a galaxy into the IGM, where $L_{UV}$
and $L_{LyC}$ describe the intrinsic luminosities of galaxies in the
$UV$-continuum and Lyman-continuum, respectively, prior to absorption
by gas or dust, and where $F_{UV}$ and $F_{LyC}$ describe the measured
fluxes of galaxies at $UV$ continuum and Lyman-continuum wavelengths,
after correction for IGM absorption.

Current observational estimates of the escape fraction at $z\sim2$-3
are substantially higher than has been measured at lower redshift,
strongly pointing towards the evolution in this escape fraction with
cosmic time from higher fractions at interemediate-to-high redshift to
very low fractions at $z\sim0$-1 \citep{BS10}.  The apparent evolution
strongly correlates with the evolution apparent in the escape fraction
of Ly$\alpha$ photons \citep{MH11} and the H I covering factor, as
inferred from UV absorption lines \citep{TJ13}.  Evolution in the
escape fraction had also been speculated to help match the Thomson
optical depths measured from the WMAP observations \citep{FH12,MK12}.

In addition to the direct constraints on the escape fraction from a
measurement of the Lyman-continuum fluxes for distant galaxies, there
are several other promising ways of constraining it.  One method makes
use of the constraints on the density of ionizing photons from studies
of the Ly$\alpha$ forest.  By modeling the observations of the
Ly$\alpha$ forest, one can quantify the photoionization rate $\Gamma
(z)$ and the mean free path for $UV$ photons $\lambda_{mfp}$.  Then,
using the proportionality $\Gamma (z) \propto \epsilon \lambda_{mfp}$
and comparing this emissivity with the luminosity density in the $UV$
continuum, one can obtain an average constraint on the escape fraction
of the total galaxy population.  The precise value of the escape
fraction one infers depends somewhat on how faint one assumes that
galaxy LF extends.  Using this technique, \cite{MK12} infer an escape
fraction of $\sim$4\% at $z\sim4$ based on the published
photo-ionization rates $\Gamma (z)$ and mean free paths from
\cite{CF08} and \cite{AS10} and the published LF results of
\cite{RB07} integrated to $-$10 AB mag.  Assuming a minimal escape of
Lyman-continuum photons from most luminous galaxies ($M_{UV,AB}<-19$),
the implied escape fraction would be $\sim$10\%.

Alternate constraints on the escape fraction of galaxies come from the
study of Gamma Ray Bursts \citep{HC07} in $z\sim2$-4 galaxies.
Gamma-ray bursts are potentially ideal for constraining the
Lyman-continuum escape fraction, if these bursts show a similar
spatial distribution in star-forming galaxies as the hot stars
producing ionizing photons.  The presence or absence of one optical
depth of neutral hydrogen absorption can be immediately seen from
optical follow-up spectrocopy of the bursts.  Of the 28 GRB systems
examined by \cite{HC07} where the underlying column density of atomic
hydrogen has been measured, only 1 shows a relative absence of neutral
hydrogen in front of the burst, resulting in $f_{esc} ^{LyC}$ fraction
estimate of $2\pm2$\% (implying a $f_{esc} = 0.04\pm0.04$ assuming
$f_{esc}^{UV} \sim 0.5$ based on the measured values of the
$UV$-continuum slope $\beta$ for faint galaxies and a Calzetti et
al.\ \citep{DC00} extinction law).

Direct studies of Lyman-continuum emission from $z\sim2$-3 galaxies
provide some evidence for galaxies at lower luminosities or with
Ly$\alpha$ emission having $\sim$2-4$\times$ higher values of the
escape fraction than for the highest luminosity galaxies
\citep{DN13,RM13}.  If true, evolution in the density of ionizing $UV$
radiation would much more closely follow the evolution in the
luminosity density of faint ($M_{UV,AB}<-19$) galaxies.  Given the
minimal evolution in the luminosity density of the faintest sources
(lowest two panels of Figure~\ref{fig:ldevol}), one might expect a
similarly slow evolution in the density of ionizing radiation with
cosmic time.  Indeed, a model with a nearly constant escape fraction
of $\sim$10\% for lower luminosity ($M_{UV,AB}>-19$) galaxies would
succeed in matching essentially all observational constraints on the
Lyman-continuum escape fraction discussed here.

\begin{figure}[h]
\includegraphics[scale=0.45]{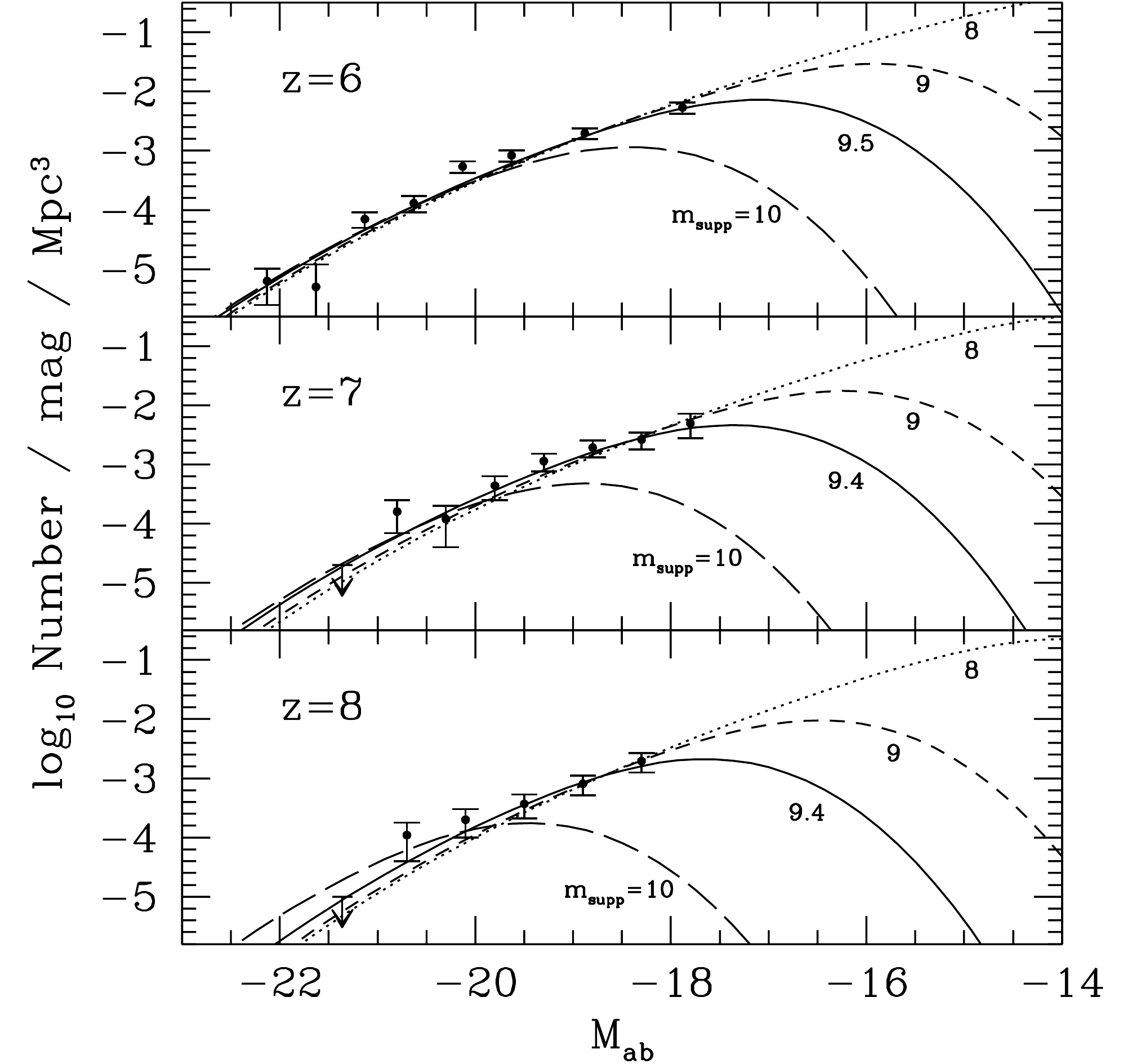}
\caption{An illustration of several model luminosity functions
  \citep{JM11} with a cut-off at the faint end (\S\ref{sec:252}).
  Results are shown relative to the LFs observed at $z\sim6$
  (\textit{top}), $z\sim7$ (\textit{middle}), and $z\sim8$
  (\textit{lowest}) panel.  Dotted, short-dashed, and long-dashed
  curves are LFs assuming $m_{\rm supp}=8$, 9, and 10, respectively,
  with the best-fit value of $L_{10}$ for each value of $m_{\rm
    supp}$.  The solid lines show results with the absolute minimum
  value of chi-square at each redshift.  The best-fit values of
  $m_{\rm supp}$ are 9.47, 9.4, and 9.42, for $z=6$, 7, and 8,
  respectively.}
\label{fig:munoz}
\end{figure}

\begin{figure}[h]
\includegraphics[scale=0.6]{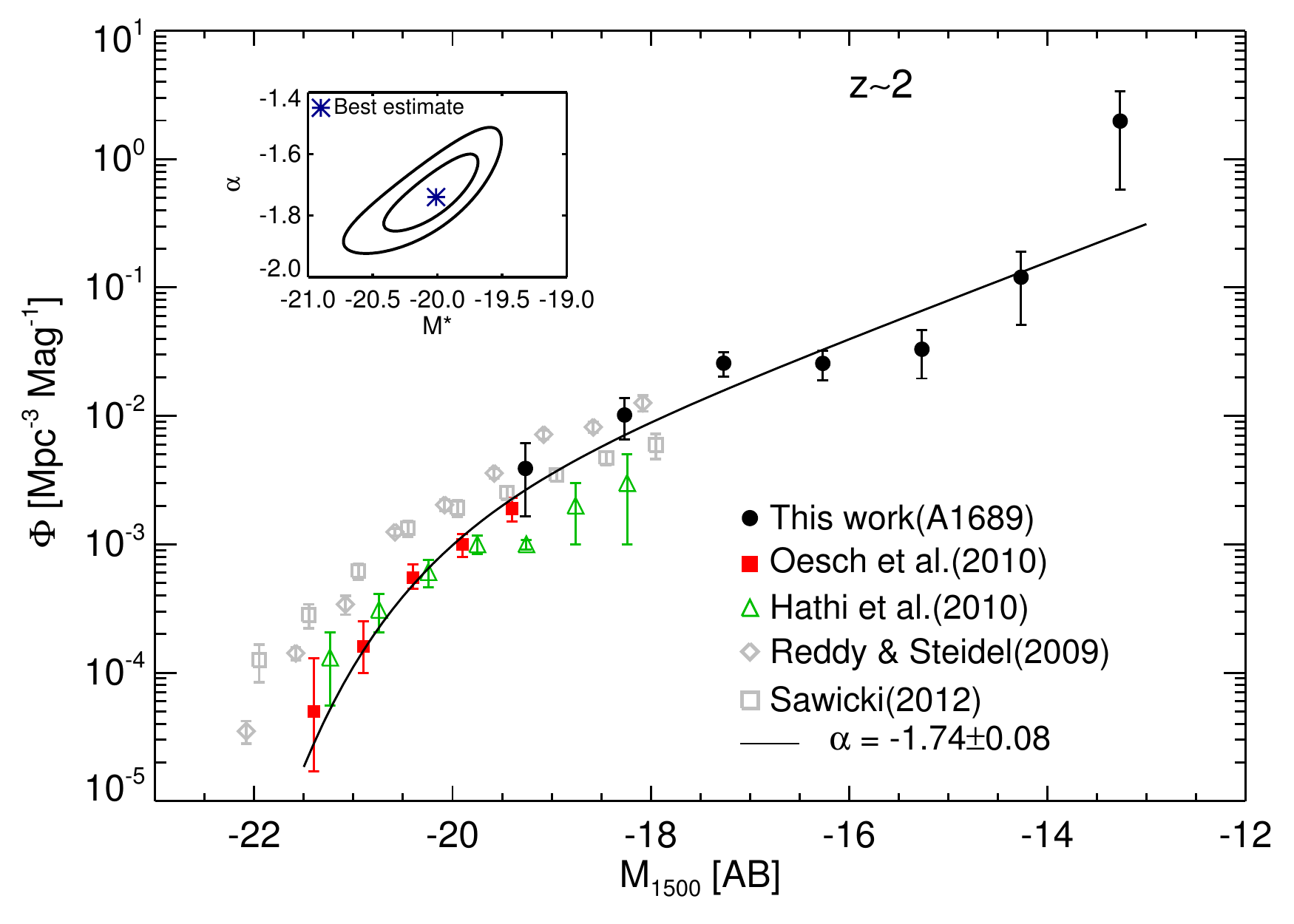}
\caption{The estimated $UV$ luminosity function at $z\sim2$ derived by
  \cite{AA14} using highly-magnified, gravitationally-lensed galaxies
  behind Abell 1689 (\textit{solid black circles}).  Also shown are
  the volume densities of more luminous galaxies at $z\sim2$ from
  wider-area probes (\textit{assorted points}:
  \cite{MS06,NR09,PO10,NH10}).  The solid line shows the $z\sim2$ LF
  derived by a combined fit to the fainter sources discovered behind
  Abell 1689 and more luminous sources found from wide-area searches.
  The volume density of faint sources discovered behind Abell 1689
  appears consistent with what is found over wide-area surveys at
  $-19$ mag, but extends to $-14$ mag with a power-law-like slope.
  This suggests that such faint galaxies plausibly exist at
  $z\sim6$-10 and can potentially reionize the universe
  (\S\ref{sec:252}).\label{fig:lf}}
\label{fig:alavi}
\end{figure}

\subsubsection{Faint End Cut-off to LF}
\label{sec:252}

If the faint-end slope $\alpha$ of the $UV$ LF is close to $-2$, then
one could potentially expect a particularly significant contribution
of galaxies at arbitrarily faint luminosities to the reionization of
the universe.  A faint-end slope of $-$2 is sufficiently steep for the
total luminosity density to be technically divergent if the integral
(Eq.~\ref{eq:intl}) is extended all the way to zero.

In reality, however, the galaxy LF cannot extend to arbitrarily low
luminosities with such a steep faint-end slope and must eventually
turn over at some luminosity.  This is due to the challenge especially
low-mass collapsed halos have in accreting or retaining significant
amounts of cool gas necessary for star formation.  There are many
compelling physical reasons to expect galaxies to only efficiently
form in halos above a certain mass \citep{JR06,JM11}.  Inefficient gas
cooling \citep{MR77}, supernovae winds \citep{MM99}, and a high UV
background \citep{MD04} are all issues that are likely to contribute
to suppressing significant star formation in very low-mass ($<10^{8}$
$M_{\odot}$) halos.  Determining at which luminosity the $UV$ LF cuts
off is clearly important for assessing the total reservoir of ionizing
radiation available to reionize the universe (e.g., see
Figure~\ref{fig:munoz}).

Ascertaining to which luminosity the UV LF extends observationally is
clearly quite challenging and will likely remain challenging for the
foreseeable future.  Searches for faint $z\sim2$ galaxies find an
abundance of galaxies to $-$14 mag \citep{AA14} and provide no
indication for an abrupt cut-off (see Figure~\ref{fig:alavi}).  Simple
reconstructions of the faint-end of the $UV$ LF using the
star-formaton histories of dwarf galaxies in the local group similarly
suggest that the $UV$ LF extends down to at least $-14$ mag and
potentially down to $-5$ mag \citep{DW14} (suggestively similar to
that found in the very high-resolution simulations of \cite{BO15}).

While one could potentially hope to observe the faintest galaxies with
gravitational lensing or with JWST, such techniques and advances in
technology only allow us to gain a factor of 10 in sensitivity over
what is state of the art at present (i.e., allowing to view galaxies
without the aid of lensing to $\sim$$-$16 mag [$\sim$$-$14 with
  lensing]).  If the faint end of the LF extends faintward of $-13$,
it is unlikely that direct probes will be successful in revealing
where the $UV$ LF ultimately cuts off.

\subsubsection{Evolution of the $UV$ LF at $z>8$}
\label{sec:253}

For galaxies at the highest redshifts $z>8$, there is much more
uncertainty regarding how the $UV$ LF evolves with cosmic time.  Most
early results on the volume density of $z\sim9$-10 galaxies
\citep{RB11A,PO12,PO13,RE13} suggested that the luminosity density of
galaxies at $z\sim9$-10 might be significantly less than what one
would derive extrapolating the $z\sim4$-8 results to $z\sim10$.

This suggested that galaxies might exhibit a slightly more rapid
evolution with time at $z>8$ than they exhibited over the redshift
interval $z=4$-8.  The rationale for the more rapid evolution observed
was unclear, but was thought to potentially arise from the halo mass
function evolution not translating into a smooth evolution of the $UV$
LF with redshift.  As \cite{PO13} demonstrate, results from a number
of independent theoretical models \citep{MT10,CL11,KF11,ST13,SG14}
predict a slightly more rapid change in the luminosity density
evolution at $z>8$ than from $z=8$ to $z=4$ (but see however the
results of the Behroozi et al. \cite{PB15} model).  Of course, another
explanation for the trend could be one of dust extinction, as the
early $z=9$-10 results show better continuity with the $z=4$-8
results, if considered after dust correction.  This idea was
implicitly first noted by \cite{RS12}, but has also been discussed in
later work \citep{RB15,SF15}.

Deeper searches over the Hubble Deep Field and parallel fields
\citep{RE13,PO13} yielded a slight deficit in the luminosity density
similar to those initially obtained by \cite{RB11A} and \cite{PO12}.
Similarly, searches for $z=9$-11 galaxies over lensing clusters also
suggested a slight deficit at $z\sim9$ relative to lower-redshift
trends (\cite{RB14B}; but see also \cite{WZ12}).  However, the
$z\sim11$ search results of Coe et al. \citep{DC13} suggested no
change in the $\rho_{UV}$ trends to $z\sim11$.

More recently, however, new $z\sim9$ and $z\sim10$ search results
\citep{AZ14,PO14B,DM14} from the Frontier Fields program \citep{DC15}
show somewhat higher volume densities than what was initially found
over the Hubble Ultra Deep Field and deep parallel fields.  At face
value, this suggests that $z\sim9$-10 galaxies might have been
underdense over the Chandra Deep Field South, and the true average
density at $z\sim9$-10 is higher and also consistent with an
extrapolation from lower redshift.  However, the Hubble Frontier
Fields program is still ongoing, and it is not yet clear whether the
evolution in $UV$ LF at $z>8$ is essentially a continuation of the
evolution from $z\sim8$ to $z\sim4$ or some acceleration is present.

\subsection{The Ionizing Photon Density Produced by Galaxies}
\label{sec:26}

Using a similar procedure to many previous analyses
\citep{PM99,MK12,BR13}, we can put together current constraints on the
evolution of the $UV$ luminosity function with an
empirically-calibrated model for the production and release of
ionizing radiation from $z>6$ galaxies.  Following \cite{BR13}, we
parameterize the total density of ionizing photons $\dot{N}_{ion} (z)$ as
\begin{equation}
\dot{N}_{ion} (z) = \rho_{UV} (z) \xi_{ion} f_{esc}
\end{equation}
where $\rho_{UV} (z)$ represents the rest-frame $UV$ luminosity
density, $\xi_{ion}$ represents the conversion factor from $UV$
luminosity to ionizing radiation, and $f_{esc}$ indicates the fraction
of ionizing radiation that escapes from galaxies after modulation by
dust and neutral hydrogen within galaxies.

\begin{figure}[h]
\hspace{1.0cm}\includegraphics[scale=0.8]{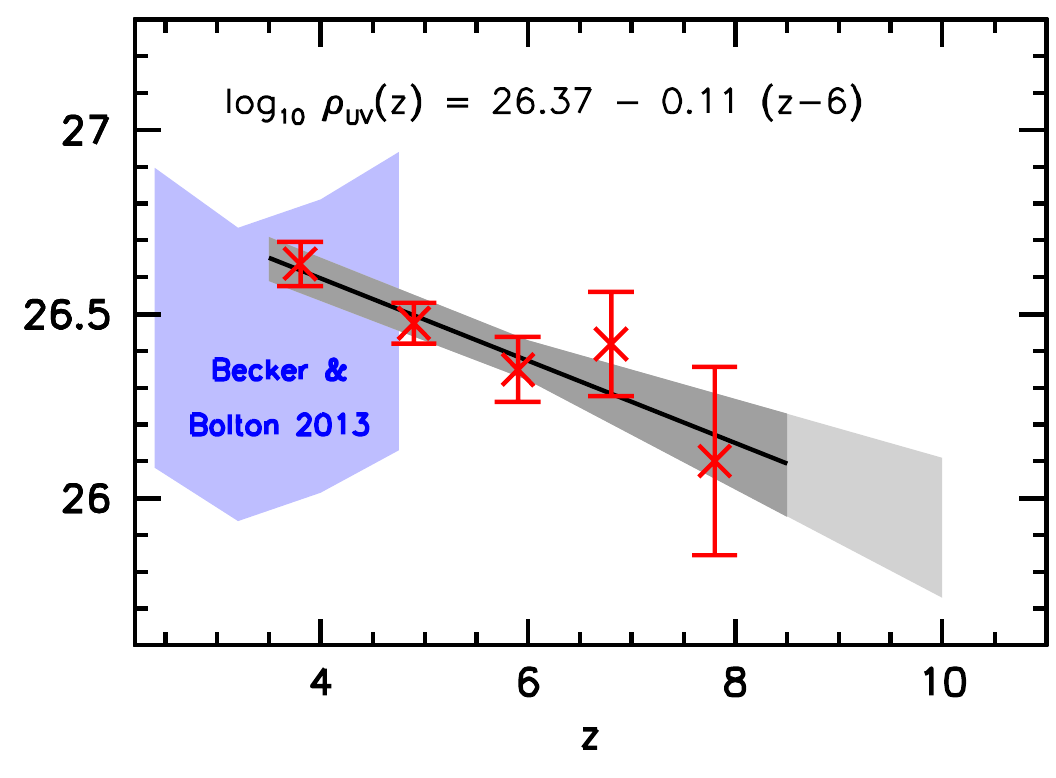}
\caption{The evolution of the $UV$ luminosity density over a larger
  range in absolute magnitude than considered in
  Figure~\ref{fig:ldevol}, i.e., $-23<M_{UV,AB}<-13$, based on the
  recent comprehensive determinations of the $z=4$-8 LFs from
  \cite{RB15} (\S\ref{sec:26}).  The points and shaded regions are as
  in Figure~\ref{fig:ldevol} and similar caveats apply.  Also shown on
  the right axis of this figure is the equivalent cosmic ionizing
  emissivity that galaxies over the specified luminosity range produce
  for the conversion factor $\log_{10} f_{esc}\xi_{ion} = 24.53$
  s$^{-1}/($ergs s$^{-1}$Hz$^{-1})$.  One recent measurement of the
  cosmic ionizing emissivity at $z\sim2$-4.75 from \cite{GB13} is also
  shown for context (\textit{light-blue-shaded area}: 68\% confidence
  region).}
\label{fig:ldevolf}
\end{figure}

In computing the cosmic ionizing emissivity $\dot{N}_{ion}(z)$ from
the $UV$ luminosity density $\rho_{UV}(z)$, we take
$\xi_{ion}=10^{25.45}$ $s^{-1} / (ergs\,s^{-1} Hz^{-1})$, consistent
with lower luminosity galaxies dominating the $UV$ luminosity density
at $z>4$ and faint galaxies having a $UV$-continuum slope $\beta$ of
$-2.2$ (\S\ref{sec:24}).  The escape fraction $f_{esc}$ is chosen so
that the universe finishes reionization at $z\sim6$, consistent with
observations, and so that the Thomson optical depth matches that seen
in the observations.  While it is not possible to motivate the value
0.11 for the escape fraction purely based on observational
considerations, the chosen value does fall within the allowed range.

The estimated ionizing emissivity from our fiducial model is presented
in Figure~\ref{fig:ldevolf}.  As should be clear by our procedure for
choosing $f_{esc}$, the overall emissivity estimated here is highly
uncertain.  The three factors that contribute to the overall estimate
could be plausibly different by 0.2 dex ($\xi_{ion}$), 0.8 dex
($f_{esc}$), and 0.4 dex ($\rho_{UV}$) from the baseline model
presented here.  While $f_{esc}$ is plausibly known to better than 0.3
dex at $z\sim3$ for $L^*$ and sub-$L^*$ galaxies, it is completely
unclear what $f_{esc}$ is for ultra-faint galaxies at $z\sim6$-10.
Uncertainties in $\rho_{UV}$ result from a lack of knowledge regarding
where the truncation of the $UV$ LF occurs.

Also shown on this same figure are constraints on the ionizing
emissivity from observations of the Ly$\alpha$ forest \citep{GB13} at
$z\sim2$-4.75 (see \cite{MK12} for earlier constraints on this
emissivity).  It is clear that the emissivity produced by the present
model is overall in excellent agreement with the observational
constraints.  One potential resolution to the slight tension between
the ionizing emissivity presented here and observational constraints
at $z\sim4$ is by accounting for the fact that a significant fraction
of the luminosity density at $z\sim4$ derives from particularly
luminous galaxies and those galaxies seem to show a lower escape
fraction \citep{DN13,RMos13}.  The ionizing emissivity produced by the
present model is also in reasonable agreement with other models in the
literature \citep[e.g.][]{BR13,BR15}.

\section{Self-Consistent Models of Reionization}
\label{sec:3}

\subsection{Standard Reionization Model Using Galaxies}
\label{sec:31}

To determine the impact of the ionizing photon production from
galaxies on the ionization state of atomic hydrogen at $z\geq5$, it is
conventional to follow the evolution of this ionization state with
cosmic time.  The simplest way to do this is by lumping the ionized
hydrogen in the $z>6$ IGM into a single quantity $Q_{HII}$, the
filling factor of ionized hydrogen, and modeling its evolution with
cosmic time with the following differential equation \citep{PM99}:
\begin{equation}\frac{dQ_{HII}}{dt}=\frac{-Q_{HII}}{<t_{rec}>}+\frac{\dot{N}_{ion} (z)}{n_H(0)}\label{eq:xe}\end{equation}
where $\dot{N}_{ion}(z)$ is the ionizing emissivity produced by the
observed population of galaxies, $n_H$ corresponds to the comoving
volume density of neutral hydrogen in the universe and $<t_{rec}>$
corresponds to the recombination time for neutral hydrogen
\begin{equation}<t_{rec}>=0.88\textrm{Gyr}\left(\frac{1+z}{7}\right)^{-3} \left(\frac{T_0}{2\times10^4 K}\right)^{-0.7} (C_{HII}/3)^{-1}\label{eq:recomb}\end{equation}
where $C_{HII}$ is the clumping factor of neutral hydrogen
$(<n_{HII}^2>/<n_{HII}>^2)$ and $T_0$ is the IGM temperature at mean
density.  The temperature $T_0$ is taken to be $2\times10^4$ K to
account for the heating of the IGM due to the reionization process
itself \citep{LH03}.  The above expression for the recombination time
$t_{rec}$ has been updated from the expression given in \cite{MK12} to
reflect the new results from Planck \citep{PL15}, where $H_0 =
67.51\pm0.64$, $\Omega_{\Lambda}=0.6879\pm0.0087$,
$\Omega_{m}=0.3121\pm0.0087$, $\Omega_{b} h^2 = 0.02230\pm0.00014$,
and $\tau=0.066\pm0.016$.  We develop a similar set of equations in
Chapter XXX.

The whole exercise of self-consistently following the evolution of
$Q_{HII}$ is a valuable one.  Importantly, it allows us to test
whether the ionizing emissivity the observed galaxy population
plausibly produces can self-consistently satisfy a variety of
different constraints on the ionization state of the universe at
$z\sim6$-10.  The most important of these constraints are the redshift
at which reionization is complete, what the ionization state of the
universe is at $z\sim7$-8 where Ly$\alpha$ emitters and galaxies can
be examined, and the Thomson optical depths $\tau$ measured from
probes of the CMB.

There are many examples of similar analyses in the literature
\cite[e.g.,
][]{TC05,JB07,PO09,FH12,RB12A,MK12,MA12,BR13,ZC14,TC14,BR15}.  Several
of the first of these analyses to discuss many of the most important
constraints discussed above include
\cite{JB07,PO09,FH12,RB12A,MK12,MA12}.  Arguably the most
sophisticated and well-developed of these include
\cite{FH12,MK12,BR13}.  However, in appreciating the insight and value
provided by such analyses, it is important to realize such analyses
ignore one important effect: the most significant sinks for ionizing
sources almost certainly lie in exactly the same regions that produce
the majority of the ionizing photons.  As these regions are denser and
therefore have higher recombination rates than the cosmic average,
this effect would cause the process of reionization to occur slower
than calculated for the simplistic models presented here \cite{ES14}.

\begin{figure}[h]
\includegraphics[scale=0.5]{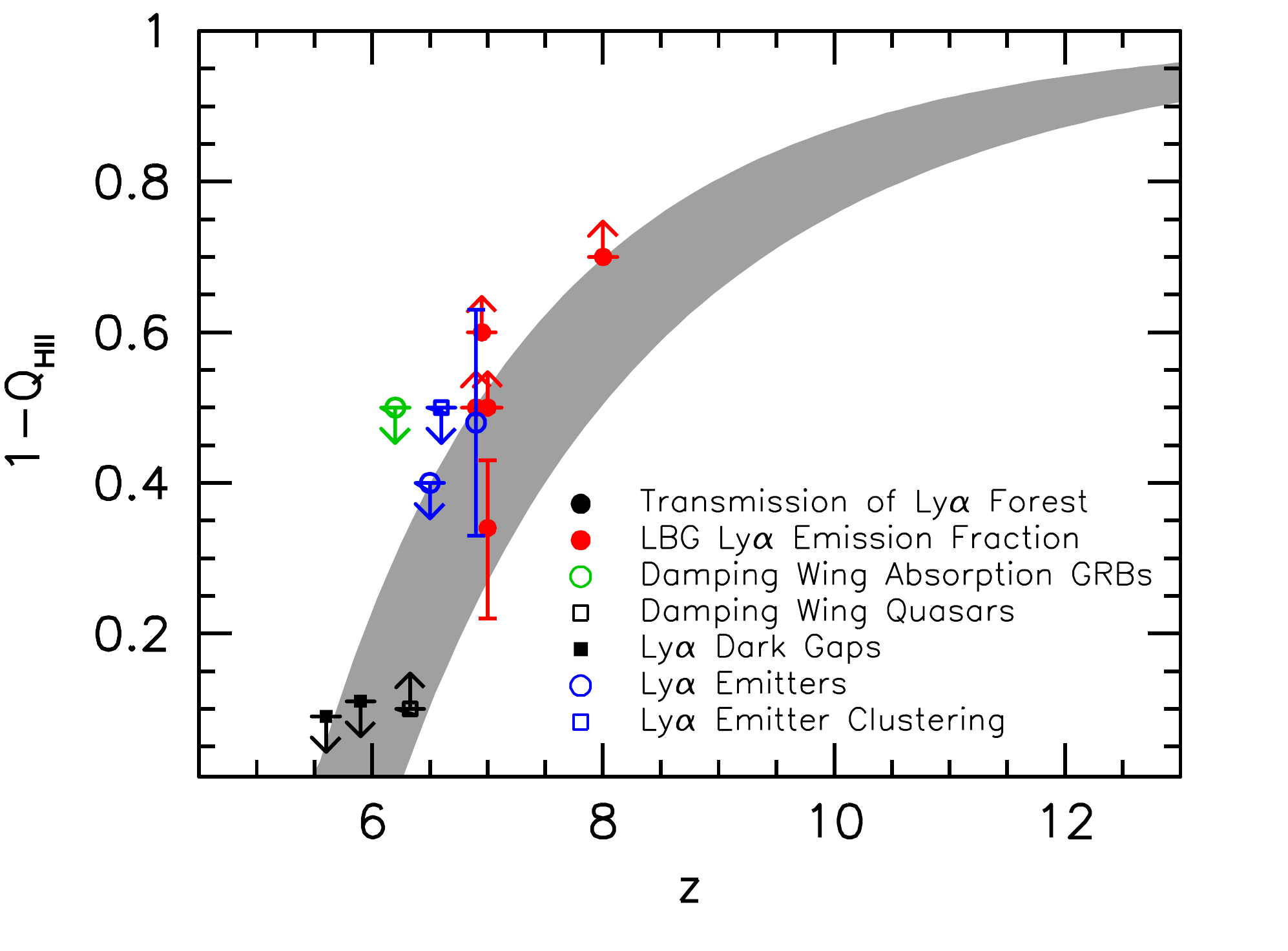}
\caption{Filling factor of neutral hydrogen ($1-Q_{HII}$: the filling
  factor of ionized hydrogen) versus redshift based on the
  evolutionary model presented in \S\ref{sec:31} based on the LF
  results from \cite{RB15}, the optical depth results from cite{PL15},
  and many other assorted constraints on the reionization of the
  universe presented in this figure.  See \cite{BR13} and \cite{BR15}
  for a detailed description of these constraints (\S\ref{sec:31}).}
\label{fig:reion}
\end{figure}

\begin{figure}[h]
\includegraphics[scale=0.6]{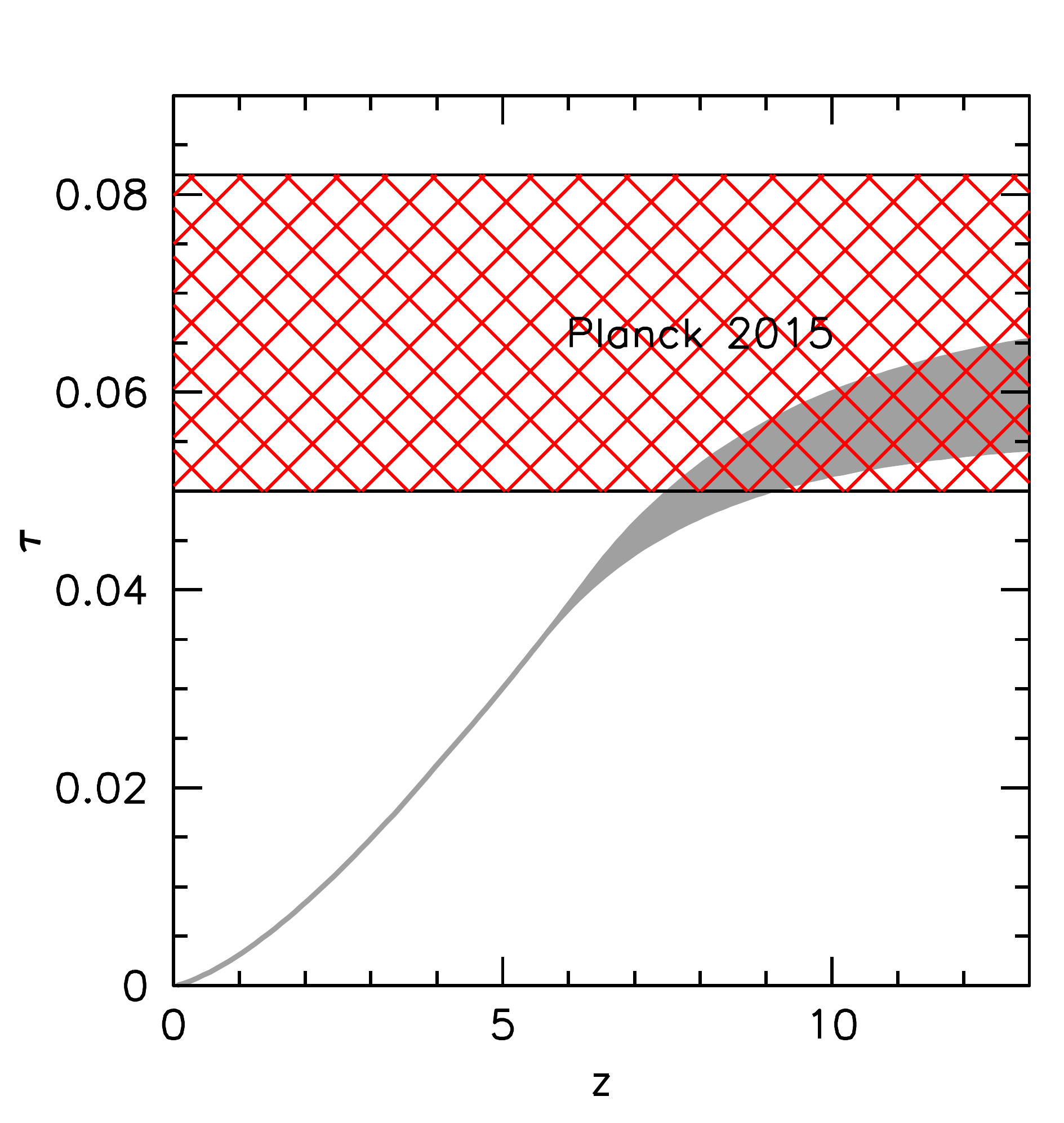}
\caption{Cumulative contribution of the ionized hydrogen and helium
  from $z=0$ to an arbitrarily high redshift (\textit{shaded grey
    contour}) from the model for the ionizing emissivity presented in
  Figure~\ref{fig:ldevolf} and assuming a clumping factor $C_{HII}$ of
  3 (\S\ref{sec:31}).  The Planck 3-year results are shown with the
  hatched redshift region $\tau=0.066\pm0.016$.}
\label{fig:tau} 
\end{figure}

Figure~\ref{fig:reion} shows how the filling factor of ionized
hydrogen evolves with redshift using the estimate we provide in
\S\ref{sec:26} of the ionizing radiation coming from galaxies
(Figure~\ref{fig:ldevolf}) and adopting a clumping factor of
$C_{HII}=3$.  The universe finishes reionization somewhere between
$z=5.5$ and $z=6.5$ in the present model, depending upon whether one
takes the ionizing emissivity to be at lower or upper end of the range
presented in Figure~\ref{fig:ldevolf}.  The calculated filling factor
$Q_{HII}$ is in plausible agreement with many of the prominent
constraints shown in Figure~\ref{fig:reion}.  The Thomson optical
depth derived from this fiducial model is also in excellent agreement
with the new Thomson optical depth constraints from Planck
$\tau=0.066\pm0.016$ (Figure~\ref{fig:tau}).

\begin{figure}[ht]
\hspace{2cm}\includegraphics[scale=0.61]{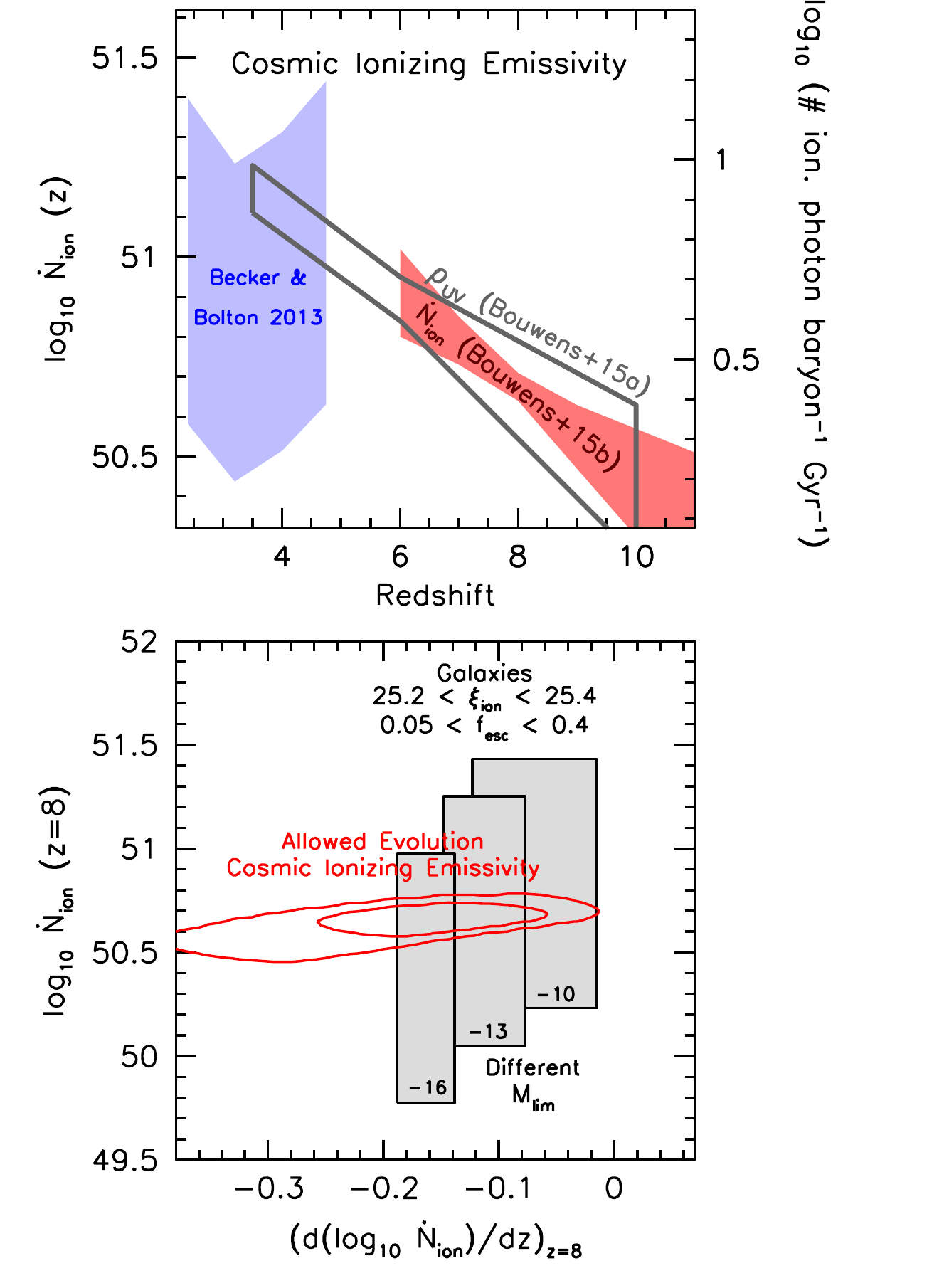}
\caption{(\textit{upper}) The evolution of the cosmic ionizing
  emissivity inferred by \cite{RB15B} (\textit{shaded red region})
  using several of the most prominent observational constraints on the
  ionization state of the universe \citep{XF06,MS14,PL15} (see
  \S\ref{sec:32}).  The vertical axis on the right-hand side gives the
  equivalent number of ionizing photons produced per Gyr per baryon in
  the universe.  The light-blue-shaded region is the emissivity at
  $z\sim2$-4.75 derived by \cite{GB13}.  The region demarcated by the
  thick grey lines shows the evolution of the $UV$ luminosity density,
  converted to an emissivity using a constant conversion factor
  $\log_{10} f_{esc}\xi_{ion} = 24.53$ s$^{-1}/($ergs
  s$^{-1}$Hz$^{-1})$.  The evolution inferred for the cosmic ionizing
  emissivity at $z>6$, i.e., $(d\log_{10}
  \dot{N}_{ion}(z)/dz)_{z=8}=-0.15_{-0.11}^{+0.09}$, is similar to the
  observed evolution in the $UV$ luminosity density for galaxies
  derived by \cite{RB15}, i.e., $\log_{10} \rho_{UV} = -0.11\pm0.04$.
  While one could consider using a slightly different set of
  constraints than what \cite{RB15B} consider to derive the evolution
  of the cosmic ionizing emissivity, this result does suggest that
  galaxies do indeed drive the reionization of the universe.
  (\textit{lower}) 68\% and 95\% likelihood contours on $\log_{10}
  \dot{N}_{ion}(z=8)$ and $(d\log_{10}\dot{N}_{ion}/dz)_{z=8}$ derived
  by \cite{RB15B} from the prominent observational constraints on the
  ionization state of the universe (\S\ref{sec:32}).  Shown with the
  grey squares are the equivalent parameters expected for galaxies
  based on the observations \citep{RB15} for a range of $f_{esc}$'s
  and $\xi_{ion}$'s for three different faint-end cut-offs to the
  luminosity function $M_{lim}$ ($-10$ mag, $-13$ mag, and $-16$ mag).}
\label{fig:ionback}
\end{figure}

\subsection{Does the Inferred Evolution in the Cosmic Ionizing Emissivity Match that Expected From Galaxies?}
\label{sec:32}

As the analysis from the previous subsection illustrates, it is clear
that one can create a self-consistent model for the reionization of
the universe using the observed galaxy population as a basis.

While such demonstrations are encouraging, they do not really address
the question of uniqueness and whether galaxies uniquely fit the
profile of those sources needed to provide the bulk of the photons for
reionizing the universe.  This is an important topic, given the large
uncertainties in the many factors that contribute to the calculation
of the ionizing emissivity for galaxies.

To address the question of uniqueness, \cite{RB15B} made use of the
some of the best constraints on the ionization state of the universe
\citep[e.g., ][]{XF06,MS14,PL15} to try to infer the evolution of the
cosmic ionizing emissivity $\dot{N}_{ion}(z)$ (see also \cite{SM15}).
In the results shown in Figure~\ref{fig:ionback}, \cite{RB15B}
presented constraints on the evolution of the emissivity towards high
redshift.  In particular, $d(\log_{10}\dot{N}_{ion})/dz$ was found to
be equal to $-0.15_{-0.11}^{+0.09}$ at $z=8$.  This is very similar to
the observed evolution in the $UV$ luminosity density of galaxies
$\rho_{UV}$ integrated to $-13$, which have an equivalent logarithmic
slope (d$\log_{10}\rho_{UV}/dz$) of $-0.11\pm0.04$ at $z=8$.  The
similar dependencies of the two quantities on redshift are illustrated
in the top panel of Figure~\ref{fig:ionback}.

To take this comparison of the galaxy luminosity density with the
required ionizing emissivity one step further, the lower panel of
Figure~\ref{fig:ionback} casts the comparison in terms of the required
emissivity $\dot{N}_{ion}(z)$ at $z=8$ and
$(d(\log_{10}\dot{N}_{ion})/dz)_{z=8}$.  Also presented in this panel
are the expected parameters for the galaxy population assuming
different escape fractions $f_{esc}$, different production rates of
Lyman-continuum photons $\xi_{ion}$, different faint-end cut-offs to
the LF.

It should be clear that one can produce the required ionizing
emissivity with a wide variety of different parameter combinations
($f_{esc}$, $\xi_{ion}$, $M_{lim}$).  Even though there are large
observational uncertainties on each of the parameters in isolation,
collectively they can be constrained quite well, using current
constraints on the ionization state of the IGM at $z>6$.\footnote{For
  example, at present, it is not really known if the escape fraction
  $f_{esc}$ of faint galaxies at $z\sim7$-8 is 0.08 or 0.3 from the
  observations.  Both possibilities could be accommodated within the
  uncertainties by making different assumptions about the faint-end
  cut-off to the LF $M_{lim}$ or to the Lyman-continuum production
  efficiency $\xi_{ion}$.}

\section{Future Prospective}
\label{sec:8}

In both the present and immediate future, progress is being made with
the Hubble Frontier Fields program \citep{DC15}.  This program is
obtaining ultra-deep images of six different galaxy clusters with the
Hubble and Spitzer Space Telescopes over a three-year period.  The
goal of this initiative is to obtain our deepest-ever views of the
distant universe, by combining the power of extraordinarily deep
exposures with Hubble and Spitzer with magnification boosts from
gravitational lensing by the galaxy clusters.  Accounting for the
$\sim$5$\times$ typical magnification factors expected from
gravitational lensing over these fields, our view from this program
will extend to sources fainter than even seen in the Hubble Ultra Deep
Field \citep{SB06,RB11,RE13,GI13}.

In total, 840 orbits of HST imaging observations are being obtained on
these clusters, reaching to $\sim$28.7 AB mag.  Half of the observing
time is being devoted to deep near-IR observations and half to deep
optical observations over these fields.  Spitzer has also invested
1000 hours of Director's Discretionary time observing these fields to
depths of 26.3-26.8 mag ($5\sigma$) in the $3.6\mu$m and $4.5\mu$m
bands, $\geq$3$\times$ deeper than the observations over the CLASH
clusters.

Longer term, progress will come when the James Webb Space Telescope
(JWST) begins operations in 2019.  The NIRCAM instrument on the JWST
will be $\sim10\times$ as efficient as the HST WFC3/IR instrument in
discovering $z\sim7$-10 galaxies and will completely revolutionize the
search for $z\geq12$ galaxies with high sensitivities to $5\mu$m.
Despite this dramatic leap in capabilities and a much improved
knowledge of the total photon output of galaxies to specific lower
luminosity limits, it seems unlikely that JWST will probe faint enough
to quantify the volume density of the lowest luminosity sources at
$z\sim7$-10.  Nevertheless, we remark that indirect detections of the
emission from these galaxies may be possible in the not too distant
future (see Chapter XXX).  Whatever the success of such experiments,
the impact of the lowest luminosity sources on the reionization of the
universe seems likely to remain an open issue for many years into the
future.

\begin{acknowledgement}
Many thanks are due to my scientific collaborators Garth Illingworth,
Pascal Oesch, and Ivo Labbe for some of the text and figures presented
here, which were developed through discussions we had for proposals we
wrote together.  Significant thanks is also due to Michael Kuhlen,
Claude Faucher-Gigu{\`e}re, and Brant Robertson who wrote recent
manuscripts which served as a guide to writing this chapter.  I also
acknowledge support from NASA grant HST-GO-11563, ERC grant HIGHZ
\#227749, and a NWO vrij competitie grant 600.065.140.11N211
\end{acknowledgement}

\end{document}